\documentclass[aps,prb,twocolumn,floatfix,superscriptaddress,showpacs,reprint,
longbilbiography]{revtex4-2}
\usepackage{amsmath, amssymb, amsthm, amsfonts,braket, mathtools,arydshln}
\usepackage{bm} 
\usepackage{graphicx, subfigure}
\usepackage[table]{xcolor}
\usepackage{times}
\usepackage{newtxtext}         
\usepackage{tikz}
\usetikzlibrary{matrix, calc}

\DeclarePairedDelimiter\abs{\lvert}{\rvert}

\definecolor{Blue}{HTML}{1F77B4}
\definecolor{Red}{HTML}{D62728}
\definecolor{Green}{HTML}{2CA02C}

\usepackage{xfrac}
\usepackage{tikz}
\usepackage[colorlinks]{hyperref}
\hypersetup{
    colorlinks=true,
    linkcolor=blue,
    filecolor=blue,
    urlcolor=blue,
    citecolor=blue,
}

\usepackage{comment}

\makeatletter
\def\maketitle{
  \@author@finish
  \title@column\titleblock@produce
  \suppressfloats[t]}
\makeatother

\newcommand{\Rnum}[1]{\uppercase\expandafter{\romannumeral#1}}

\newcommand{\maintitle}{Operational meaning of Markov gap in tripartite entanglement of quantum dynamics}

\begin{document}

\title{\maintitle}
\author{Zongsheng Zhou}
\affiliation{Institute of Physics, Chinese Academy of Sciences, Beijing 100190, China}

\author{Riqiang Zhang}
\affiliation{Institute of Physics, Chinese Academy of Sciences, Beijing 100190, China}
\affiliation{School of Physical Sciences, University of Chinese Academy of Sciences, Beijing 100049, China }

\author{Yu-Xiang Zhang}
\email{iyxz@iphy.ac.cn}
\affiliation{Institute of Physics, Chinese Academy of Sciences, Beijing 100190, China}
\affiliation{School of Physical Sciences, University of Chinese Academy of Sciences, Beijing 100049, China }

\date{\today}

\begin{abstract}
We investigate how irreducible multipartite entanglement, a long-range correlation by nature, can emerge from short-range dynamics far from equilibrium. Focusing on the Markov gap as a probe of irreducible tripartite entanglement (IrTE) in free-fermion chains, we uncover qualitatively distinct dynamical behaviors:
the Markov gap grows either quasi-linearly or in staircase-like jumps depending on the initial state. We also propose attainable upper and lower bounds for the onset time of IrTE based on the Lieb-Robinson bound.  Strikingly, the Markov gap saturates to a volume-law value on a timescale  $t\sim\! L^2$, much slower than the ballistic spreading of bipartite correlations. 
To understand what information about the wavefunctions is 
revealed by the Markov gap calculation,
we introduce the concept of essential tripartite fermion (ETF)
and an associated tripartite null matrix.
The value of Markov gap closely tracks the number of small singular values of this
tripartite null matrix, yielding a transparent, operational physical 
interpretation of the measure. We further demonstrate that several dynamical signatures persist in the interacting XXZ chain.
\end{abstract}

\maketitle

\emph{Introduction.}--Entanglement growth following a global quantum quench is one of the central themes in non-equilibrium many-body dynamics~\cite{Calabrese2005, Calabrese2006PRL,Calabrese2007JSM, Fagotti2008PRA,Casini2016JHEP,Mitra2018Ann}. In short-range interacting systems, entanglement builds up immediately along the border between two initially uncorrelated subsystems, 
and spreads out ballistically within a Lieb–Robinson light cone~\cite{Lieb1972CMP, Cheneau2012Nat}. 
This picture is anchored to the quasi-particle interpretation~\cite{Calabrese2005,Calabrese2007JSM, Alba2017PNAS,Bianchi2018JHEP,Vincenzo2018SciP}. Yet the very locality that governs this picture raises a largely unexplored question: How is the multipartite entanglement assembled from local interactions, especially when the distant parties share no contiguous points? To be specific, in this Letter we study \emph{irreducible tripartite entanglement} (IrTE), where by the word ``irreducible'' we exclude the bipartite entanglement within a tripartite many-body state~\cite{Linden2002PRL,ZhouDL2008PRL,Navascues2020PRL}.  IrTE is more stringent than genuine multipartite entanglement~\cite{Otfried2009PhysR, Horodecki2009RMP,Huber2013PRA,Michael2013Science,Wilde2015ProcA,Beckey2021PRL,XieSongbo2021PRL}, which, although more widely studied, does not preclude pairwise entanglement.

Recently, IrTE has attracted growing interest in both many-body physics~\cite{Nezami:2020aa,Siva:2022aa,KanePRX2022,Liu:2022aa,Liu:2024ab,BergPRL2025,ZouYijian2021PRL, LiuKe2026Arxiv} and high energy physics~\cite{Akers2020JHEP,Hayden2021JHEP,Balasubramanian2025JHEP,Iizuka2025JHEP,SharmaPRD2022,Chris2024,iizuka2025,iizuka2025b,SimonPRL2026}.  For quench dynamics, one of the authors and colleague numerically showed that the tripartite Greenberger-Horne-Zeilinger (GHZ) entanglement in random Clifford circuits emerges through a dynamical phase transition and saturates at a finite value~\cite{Xu2025prb}. As GHZ is not the only inequivalent type of IrTE~\cite{DurW2000PRA}, here we consider IrTE quantified by the Markov gap~\cite{Akers2020JHEP, Hayden2021JHEP, Balasubramanian2025JHEP, Iizuka2025JHEP}, which vanishes for GHZ states~\cite{ZouYijian2021PRL}. However, any measure of multipartite entanglement  is inherently biased. Different measures illuminate distinct aspects, and the ranking of states under one measure is generally not preserved by another~\cite{Walter2016BookCh14,Horodecki2024Arxiv,Srivastava2024B,Mamengru2025FR}. This bias forces a fundamental question---giving its involved definition, if a nonzero Markov gap merely certifies the presence of IrTE, how should we interpret the specific value it assigns?

To address these questions, here we study the growth of IrTE in free fermion chains~\cite{Bravyi2005QIC,Peschel2009JPA,Jacopo2022SciPostPhys}, which provide a minimal setting while already exhibiting remarkably rich behavior. We identify the following main features in the quench dynamics of the Markov gap: 
(1) The Markov gap emerges after a threshold time $t_*$, for which we find attainable upper and lower bounds based on the Lieb-Robinson bound, and we identify how different initial states realize distinct onset times within this window.
(2) At the early stage, the Markov gap may display linear grow  in time (despite of a kink), or staircase-like jumps, depending on the initial conditions.
(3) The Markov gap shows a volume-law scaling. 
(4) For structured initial states (to be specified), the Markov gap saturates in a time $t{\sim} L^2$ 
where $L$ denotes the system size, far beyond the ballistic $t\sim L$ saturation characteristic of bipartite entanglement after quantum quenches~\cite{Calabrese2005, Fagotti2008PRA}.
The threshold time $t_*$ and volume-law saturation persist in interacting fermions, as confirmed by matrix product state~(MPS)~\cite{Schollwock2011Aphys,Orus2014APhys,Cirac2021RMP} simulations of the XXZ spin chain.
 
\begin{figure*}[tb]
  \includegraphics[width=0.95\textwidth]{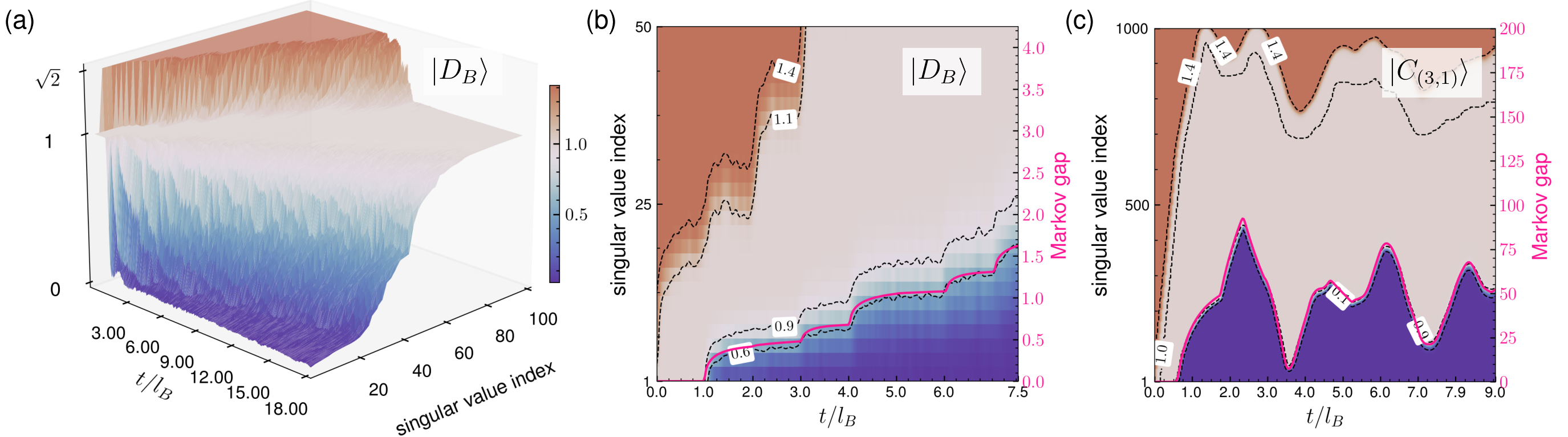}
 \caption{(a) The singular value spectrum (only the smallest 100 singular values are
 displayed) of the null matrix $\mathcal{N}_3(t)$ for the
 free fermion chain initialized in $\ket{D_B}$. The singular values cluster around 0, 1 and $\sqrt{2}$, forming three plateaus, with cliffs between them. (b)Top view of (a), with four contours (0.6, 0.9, 1.1, 1.4). 
The pink curve shows the evolution of the Markov gap, 
with values referenced to the right vertical axis. (c) Same with (b) but 
for crystal initial state $\ket{C_{(3,1)}}$ with
four contours (0.1, 0.9, 1.1, 1.4). The cliffs are steeper than (b). In (b) and (c)
we see agreement between the pink curves and the boundary of singular values around 0.
}\label{Fig3D}
\end{figure*}

To elucidate the physical meaning of the Markov gap, we introduce the concept of \emph{essential tripartite fermions} (ETFs) by exploiting equivalent representations of any free fermion state. ETFs are then characterized by an evolving null matrix $\mathcal{N}_3(t)$ (defined below). 
As illustrated in Fig.~\ref{Fig3D}(a), the singular value spectrum of $\mathcal{N}_3$ clusters
around 0, 1 and $\sqrt{2}$. Surprisingly, we find that the evolution of Markov gap closely follows
the number of singular values concentrated near 0, as demonstrated in Fig.~\ref{Fig3D}(b) and (c) for two initial states giving rise to either staircase-like jumps or a linear growth, respectively. This remarkable agreement reveals a direct connection between the Markov gap and the formation of ETFs, linking Gaussian IrTE to ETFs and their interplay with an underlying bipartite Fermi sea.

\emph{Systems and the Markov gap}---We consider the free fermion chain governed by the Hamiltonian
\begin{equation}
    H_1 = \frac{1}{2} \sum_{i} \left( c_i^\dagger c_{i+1}^{\phantom{\dagger}} + c_{i+1}^\dagger c_i^{\phantom{\dagger}} \right),
    \label{Ham:A}
\end{equation}
where $c^\dagger_i$ and $c_i^{\phantom{\dagger}}$ denote the creation and annihilation operators 
of the on-site fermion, respectively.
To account for the dependence on the initial conditions, we consider evolution from two initial conditions
(1) \textit{Crystal states} $|C_{(p, q)}\rangle = \prod_{i=1}^{L/p} \bigl(\prod_{j=0}^{q-1} c_{pi-j}^\dagger\bigr) |0\rangle$, where $L$ is the size of the chain and $|0\rangle$ denotes the empty chain. We choose $\ket{C_{(3,1)}}$ as a representative example in the main text. 
(2) \textit{Domain-filled states} $|D_B\rangle = \prod_{i \in B} c_i^\dagger |0\rangle$, where $B$ is fully occupied whereas $A$ and $C$ are empty.
In Supplemental Material, we present results for two more initial states: (3) \textit{Dimer state} $|D\rangle = 2^{-L/4} \prod_{i} \bigl(c_{2i-1}^\dagger + c_{2i}^\dagger\bigr) |0\rangle$, which is a product of  nearest-neighbor entangled pairs and (4) \textit{Random filling state} $|R_\delta\rangle$, in which sites are occupied randomly with a filling fraction $\delta$.  These two states shed light on the influence of short-range entanglement and disorder, respectively. In the main text, we employ open boundary conditions and equally divide the 1D system into three contiguous regions denoted by $A, B$ and $C$. 
Periodic boundary conditions and alternative partitions are discussed in the Supplemental Material~\cite{sp}.

To introduce the Markov gap, we first construct the square root of the reduced density matrix of the joint system $AB$, $\sqrt{\rho_{AB}}$. Its canonical purification  $\ket{\sqrt{\rho_{AB}}}$ is defined 
by mapping the operator basis element of the joint system
$\ket{a}_A\bra{a'}\otimes\ket{b}_B\bra{b'}$ to the state $\ket{a,b,a',b'}$,
thereby embedding $\ket{\sqrt{\rho_{AB}}}$ into the doubled Hilbert space $\mathcal{H}_A \otimes \mathcal{H}_B \otimes \mathcal{H}_{A^*} \otimes \mathcal{H}_{B^*}$. Then, the von Neumann entropy of the reduced density matrix $\rho_{AA^*}=\mathrm{tr}_{BB^*}(\ket{\sqrt{\rho_{AB}}}\bra{\sqrt{\rho_{AB}}})$ defines the so-called reflected entropy $S_R(A{:}B)$~\cite{Bueno2020JHEP,Dutta2021JHEP}. The Markov gap for $AB$ is then $\mathcal{M}=S_R(A{:}B)-I(A{:}B)$, where $I(A{:}B)$ is the quantum mutual 
information~\cite{Groisman2005PRA,Nielsen2010}. 
The Markov gaps for AC and BC can be defined similarly. Although they
do not equal exactly, each of them quantifies IrTE~\cite{Akers2020JHEP, Hayden2021JHEP, ZouYijian2021PRL}. Hence, we choose the Markov gap for $AB$ by default. Numerically, for fermionic Gaussian states, the above construction admits an efficient formulation in terms of the two-point correlation matrix $\mathcal{C}_{xy}(t) = \bra{\psi(t)} c_x^\dagger c_y^{\phantom{\dagger}} \ket{\psi(t)}$~\cite{Bravyi2005QIC,Peschel2009JPA,Jacopo2022SciPostPhys}, which we use throughout this work~\cite[\ref{Supp:CM}]{sp}.

\begin{figure}
    \centering
    \includegraphics[width=\linewidth]{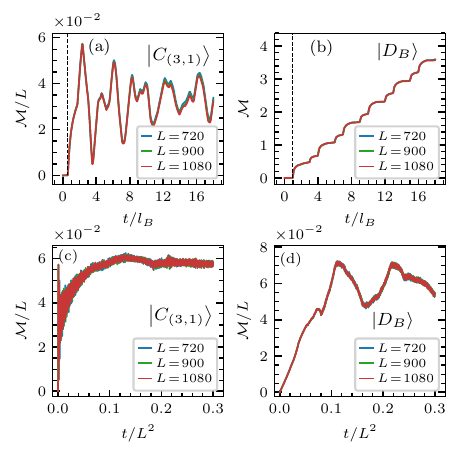}
    \caption{Upper panels $\left(\mathrm{a}\right)$ and 
    $\left(\mathrm{b}\right)$ show the early time dynamics of states
    $\left|C(3,1)\right\rangle$, and $\left|D_B\right\rangle$ for different system sizes, respectively. For each of these states, there exist a threshold, marked by a vertical dashed line, below which the Markov gap is absent. This threshold is determined by the effective light cone speed $v$.  Here, the time is rescaled by the length of subsystem $B$, under this scaling, the position of the threshold is given by $1/v$. 
    The vertical of state $\left|C(3,1)]\right\rangle$ is rescaled by $L$ to achieve data collapse. The lower panels $\left(\mathrm{c}\right)$ and $\left(d\right)$ display long time behavior of the Markov gap. The results for different system sizes collapse onto a single curve when time is scaled by $L^2$ and Markov gap is rescaled by $L$.}
    \label{Fig2}
\end{figure}

\emph{Early-time behavior.}---Hamiltonian Eq.~\eqref{Ham:A}
has the maximal group velocity $v_{\max}=1$, which sets the Lieb-Robinson
speed. Fig.~\ref{Fig2}(a) and (b) show the time evolution of Markov gap against $t/l_B$ ($l_B=L/3$ is the size of $B$) for the crystal state  $|C_{(3,1)}\rangle$ and the domain-filled state $\ket{D_B}$, respectively. 
In Fig.~\ref{Fig2}(a), $\mathcal{M}(t)$ grows approximately linearly in time despite the presence of a pronounced kink, which may signal a crossover between two different dynamical regimes. Such a feature is absent in the evolution of bipartite entanglement. Following a rapid rise to a large peak, the Markov gap undergoes substantial oscillations. The successful collapse of $\mathcal{M}/L$ for different system sizes indicates a volume-law scaling of the Markov gap.
In contrast, the domain-filled state $\ket{D_B}$ shows a periodic staircase-like growth, as shown in Fig.~\ref{Fig2}(b). The data for different system sizes collapse without any rescaling of the vertical axis, demonstrating that the height of each step is an $O(1)$ quantity independent of $L$. Remarkably, each jump is accompanied by a phase-transition like behavior~\cite{sp}.

On the other hand, the data collapses in Fig.~\ref{Fig2} as a function of rescaled time $t/l_B$ suggests a ballistic propagation of information. A manifestation is the existence of a threshold time $t_{*} \!=\! t/(l_Bv_{\mathrm{eff}})$, as indicated by the vertical dashed lines in Fig.~\ref{Fig2}, prior to which the Markov gap remains strictly zero. The value of $t_*$ takes different values for different 
initial states, implying that the onset speed of IrTE is not universal. In the following, we propose general lower and upper bounds for $t_*$.
  
\emph{Bounding $t_*$}---Consider an interaction picture defined by the free Hamiltonian $H_0=H_A+H_B+H_C$, where $H_A$ contains all terms of the Hamiltonian entirely within subsystem $A$, and similarly for $H_{B/C}$. The remaining part of the Hamiltonian consists of the boundary couplings $O_{AB}$ and $O_{BC}$, acting across the $A$--$B$ and $B$--$C$ interfaces, respectively. Then, the unitary evolution reads
\begin{equation}
\label{eq:U}
    U(t)=\hat{T}\exp\bigg\{\!-i\int_0^t\! d\tau 
    \left[O_{AB}(\tau)+O_{BC}(\tau) \right]\!\bigg\},
\end{equation}
where $\hat{T}$ denotes the time-ordering operator,  $O_{AB}(\tau)=e^{iH_0 \tau}O_{AB}e^{-iH_0\tau}$ and $O_{BC}(\tau)=e^{iH_0 \tau}O_{BC}e^{-iH_0\tau}$.
Eq.~\eqref{eq:U} generates identical entanglement as the Schr\"{o}dinger picture evolution, since they differ only by the local unitary
$e^{-iH_0 t}=e^{-iH_At}\otimes e^{-iH_Bt}\otimes e^{-iH_C t}$. 

From the Lieb-Robinson bound, the spatial extent of $O_{AB}(\tau)$ and $O_{BC}(\tau)$ expand into subsystem $B$ and propagate toward each other. Before the two light cones overlap, namely for $t< l_B/2v_{\max}$, the Hilbert space of $B$ can be decomposed as $\mathcal{H}_B = \mathcal{H}_{B_L}\otimes\mathcal{H}_{B_R}$ such that
$U(t)=U_{AB}(t)\otimes U_{BC}(t)$, where $U_{AB}=\hat{T}e^{-i\int_0^t O_{AB}(\tau)d\tau}$ and  $U_{BC}=\hat{T}e^{-i\int_0^t O_{BC}(\tau)d\tau}$. Meanwhile, the initial state $\ket{\Psi_0}$ is a product state under this decomposition. As a result, the evolved state $U(t)\ket{\Psi_0}$ contains only pairwise entanglement across $A$--$B$ and $B$--$C$ interfaces.  Thus, IrTE cannot be generated for  $t < l_B/2v_{\max}$, implying the lower bond $t_*\ge l_B/2v_{\max}$. In the Supplemental Material we will show that this bound is attained  both by the dimer state $\ket{D}$ and the random filling states $\ket{R_\delta}$.

For free-fermion systems, locality of the single-particle propagator implied by the Lieb--Robinson bound leads to an approximate decomposition $U(t)=U_{AB}(t)\otimes U_{BC}(t)$ up to $t\le l_B/v_{\max}$. Now the factorization is defined with respect to splitting $\mathcal{H}_B$ into roughly a ``right moving'' part 
and a ``left-moving'' part, where $O_{AB}(\tau)$ and $O_{BC}(\tau)$ belongs, respectively. However, the initial states, while having a product form of localized fermions, are generally entangled states with respect to  such ``de-localized'' decomposition of $\mathcal{H}_B$. A state can be represented in the form of $\ket{\Psi_0}=\sum_{i} \alpha_i\ket{\phi_{i}^{(1)}}
\ket{\phi_i^{(2)}}$ by Schmidt decomposition, hence $\ket{\Psi(t)}=\sum_{i} \alpha_i
U_{AB}(t)\ket{\phi_{i}^{(1)}}\otimes U_{BC}(t) \ket{\phi_i^{(2)}}$, which is likely IrTE. Therefore, $l_{B}/v_{\max}$ is a generic upper bound for $t_*$. Interestingly, this upper bound is attained by the domain-filled state $\ket{D_B}$. As a completely filled state in $\mathcal{H}_B$, it is invariant under basis rotations and remains a product state in above de-localized decomposition.

The threshold time $t_*$ of the crystal states are found between the upper and the lower bounds. Due to their structured pattern, many-particle interference
leads to slower effective Lieb-Robinson speed.  For $\ket{C_{(3,1)}}$,  we find that $t_*=l_B/2v_{\text{eff}}$ with $v_{\text{eff}}=\sqrt{3}/2v_{\max}$, which is consistent with the effective light cone derived in Ref.~\cite{SinghaRoy2022PRB} and the observed onset of Markov gap in Ref.~\cite{BerthiereArxiv2024}.

\emph{Long-time behavior}---To know where does the Markov gap saturates
and when the equilibrium is achieved, 
we show the long-time behavior of the 
rescaled Markov gap $\mathcal{M}(t)/L$ in Fig.~\ref{Fig2}(c,d). 
The data collapse onto $\mathcal{M}(t)/L$ immediately implies a
volume-law of the Markov gap. Moreover, although the 
short-time fine structures have a time scale $t\sim L$, 
the long-time behavior shows a time scale of $t\sim L^2$ as demonstrated by
the horizontal axes used for data collapse. 
Such time scale is best understood in the case of $\ket{D_B}$:
$t\sim L^2$ is necessary provided the volume-law of saturation, 
the $O(1)$ height of each stair, and the $t\sim L$.  This is far beyond the time scale of entanglement scrambling (which is implicitly specified for bipartite entanglement), indicating a
dynamical hierarchy in entanglement growth. The next question is
thus what structure of the wavefunction is developing but invisible for (bipartite) entanglement scrambling. 

\emph{Tripartite entanglement structure}---While our numerical results are obtained through the two-point correlation matrix, the following analysis is based on an observation about the wavefunctions: Any $n$-particle state can be expressed as $\ket{\Psi}\!=\!\prod_{i=1}^{n} d_i^\dagger\ket{0}$, where the occupied fermionic modes ($d_i^\dagger$) are mutually orthogonal. Since any unitary transformation among the occupied modes leaves the many-body state invariant, it can be equally represented as $\prod_{q=1}^n f_q^\dagger\ket{0}$, with $f^\dagger_q\!=\!\sum_i T_{q,i}d^\dagger_i$, where $T$ is an $n\times n$ unitary matrix. Thus, we can exploit this freedom to localize each $f$-fermion onto as few subsystems as possible under the tripartition $A\cup B\cup C$. By expanding every $f$-fermion onto the on-site fermions $f^\dagger_q=\sum_{i=1}^{L}f_{q,i}c_i^\dagger
\equiv \vec{f}_q\cdot\vec{c}^\dagger$, the modes can be classified into three families according to the support of $\vec {f}_q$:
(1) Single-party fermions (irrelevant to entanglement).
(2) Bipartite fermions, whose support spans two subsystems. Let $F_{AB}$ denote the space spanned by those $\vec{f}$ supported on $A\cup B$ but vanishing on $C$.  Similarly, we define $F_{BC}$ and $F_{AC}$. Because fermionic modes are anti-commute, these vectors are mutually orthogonal, i.e. $F_{AB}\perp F_{BC}\perp F_{AC}$. Thus they form the bipartite subspace $V_2=F_{AB}\oplus F_{BC} \oplus F_{AC}$. 
(3) Tripartite fermions, whose vector representations have nonzero support on all the three subsystems. By optimizing over all admissible unitary transformations $T$, the number of tripartite modes can be minimized. We refer to the resulting minimal set as \emph{essential tripartite fermions} (ETFs), and denote by $V_3$ the space spanned by their vector representations.

Starting from the bipartite fermions, note that $F_{AB}\perp F_{BC}$ implies  $\pi_B(F_{AB})\perp \pi_B(F_{BC})$,
where $\pi_X$ denotes the projection onto Party--$X$. 
This means that bipartite fermions associated with $F_{AB}$ and $F_{BC}$ occupy orthogonal degrees of freedom once restricted onto $B$. As a consequence, $V_2$  is embedded in a two-times-larger subspace
\begin{equation}\label{eq:V2}
    V^{(2)}= V^{(2)}_{AB}\oplus V^{(2)}_{BC} \oplus V^{(2)}_{AC}
    \end{equation}
where $V^{(2)}_{AB}=\pi_A(F_{AB})\oplus \pi_B(F_{AB})$ 
and similarly for $V^{(2)}_{BC}$ and $V^{(2)}_{AC}$. 
Equivalently, the Hilbert space admits a tensor-product decomposition such that the state with all bipartite fermions filled can be written as $\ket{\Psi_2}=\ket{\psi_{AB}}\otimes\ket{\psi_{BC}}\otimes\ket{\psi_{AC}}$.

\begin{figure}
    \centering
    \includegraphics[width=0.8\linewidth]{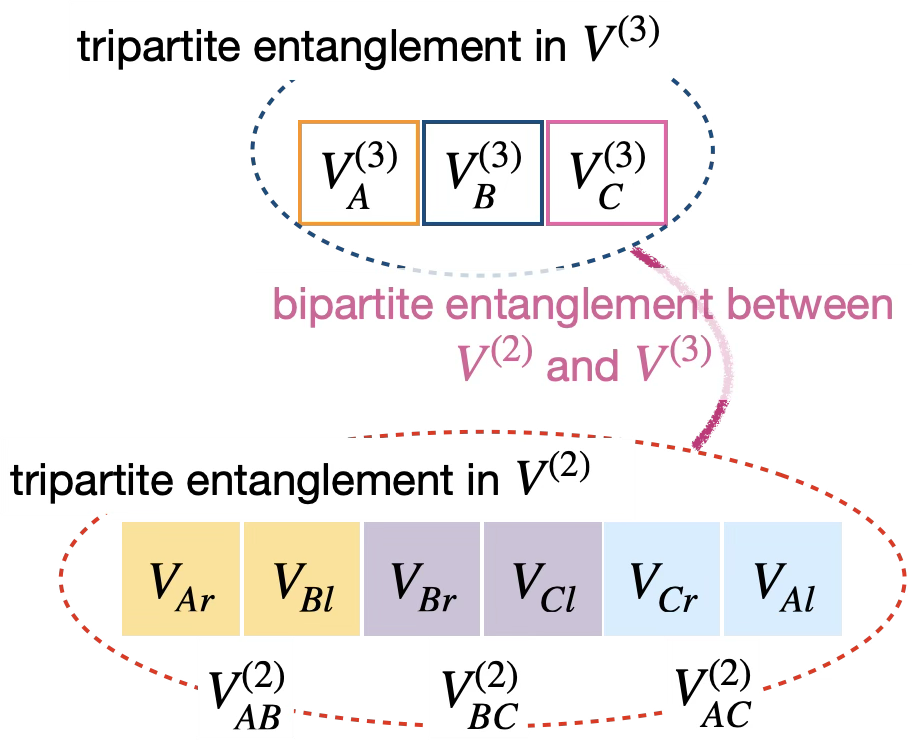}
    \caption{Schematic diagram of tripartite entanglement structure of 
    free fermion states. There are IrTE generated by ETFs in $V^{(3)}$,
    and IrTE in the ``Fermi sea'' of bipartite fermions 
    generated by the projections of the ETFs in $V^{(2)}$. There are also
    bipartite entanglement with respect to them.}
    \label{fig_structure}
\end{figure}

IrTE emerges upon filling an ETF into the bipartite sector state $\ket{\Psi_2}$. 
Crucially, $V_3\perp V_2$ does not imply $V_3\not\perp V^{(2)}$.
As a result, there is generally no tensor-product decomposition of the local Hilbert spaces under which ETF insertion yields a product state of the form $\ket{\Psi_2}\otimes\ket{\Psi_3}$.
Explicitly, consider an ETF represented by $\vec{t}\in V_3$
that satisfies $\vec{t}\not\perp V^{(2)}$. It can be expanded as 
\begin{equation}
\label{eq:t1}
    \vec{t}=\pi_{V^{(2)}}(\vec{t})+ \vec{t}'
\end{equation}
where $\vec{t}'\perp V^{(2)}$. Let $V^{(3)}_X$ denote the subspace spanned by
the projections $\pi_X(\vec{t}')$ of all such $\vec{t}'$ onto Party--$X$, and define 
$V^{(3)}=\oplus_X V_X^{(3)}$. 
Assigning a vacuum state $\ket{0}_3$ to the degrees of freedom in $V^{(3)}$, the action of filling $\vec{t}$ into $\ket{\Psi_2}$ produces
\begin{equation}
\label{eq:tpsi}    t^\dagger\ket{\Psi_2}\otimes\ket{0}_3=\ket{\Psi_2}\otimes\ket{t'}+\ket{\Psi_2^{-t}}\otimes\ket{0}_3.
\end{equation}
Here, $\ket{t'}=t'^\dagger\ket{0}_3$ has IrTE of the W-type~\cite{DurW2000PRA}:
A superposition of filling a fermion into one of all the three possible $V_X^{(3)}$.
The orthogonal complement term reads
\begin{equation}
\label{eq:Wt}
\begin{split}
   |\Psi_2^{-t}\rangle =& \left|\psi_{AB}^{-\alpha}\right\rangle \left|\psi_{BC}\right\rangle \left|\psi_{AC}\right\rangle + \left|\psi_{AB}\right\rangle \left|\psi_{BC}^{-\beta}\right\rangle \left|\psi_{AC}\right\rangle\\
    &+ \left|\psi_{AB}\right\rangle \left|\psi_{BC}\right\rangle 
    \left|\psi_{AC}^{-\gamma}\right\rangle,
\end{split}
\end{equation}
where $\ket{\psi_{AB}^{-\alpha}}$ is obtained by digging a ``hole'' in $\ket{\psi_{AB}}$, and similarly for $\ket{\psi_{BC}^{-\beta}}$
and $\ket{\psi_{AC}^{-\gamma}}$. State~\eqref{eq:Wt} also exhibits W-type IrTE, as it corresponds to an uncertainty in which of the three two-party subspaces $V_{AB}^{(2)}$, $V_{BC}^{(2)}$, or $V_{AC}^{(2)}$ contains the hole. They describe the back-action of removing weight from the bipartite sectors(see the full expressions and explanations in End Matter,
where an example of IrTE contributed by only such hole is given). The decomposition~\eqref{eq:Wt} does not respect the natural tripartite partitioning into parties $A$--$B$--$C$. When expressed in that basis, the state assumes a considerably more cumbersome form. 

State~\eqref{eq:tpsi} elucidates the  entanglement structure depicted in Fig.~\ref{fig_structure} and summarized in its caption. In particular, $\ket{\Psi_2}$ is reminiscent of the ``Fermi sea'' for IrTE because it generates IrTE when ETFs have nonzero projections in $V^{(2)}$. This raises the question of whether the observed Markov-gap results are truly measure-independent or merely definitional. Equivalently, one may ask how faithfully the Markov-gap curves encode the structure of the underlying ETFs.

We characterize the ETFs as follows. We write $\ket{\Psi(t)}$ using the 
lattice-site fermions $\ket{\Psi(t)}=\prod_{q}\left[\sum_{j}\Gamma_{q,j}(t)c_j^\dagger\right]\ket{0}$, 
where $\Gamma(t)$ defines an $n\times L$ matrix. Its columns are split
into three sub-matrices $\Gamma=\left[\Gamma_A,\Gamma_B,\Gamma_C\right]$. 
Denote by $N_X$ the matrix whose columns constitute an orthonormal basis for the kernel of $\Gamma_X^\text{T}$ ($\text{T}$ denotes transpose). 
Physically, the rows of, e.g., $N^\text{T}_B\Gamma$,  give the bipartite fermions in $F_{AC}$ and single-party 
fermions restricted in $A$ or $C$.
Then, we collect them into
a tripartite null matrix $\mathcal{N}_3 = \left[N_A, 
N_B, N_C\right]$ and identify its singular value spectrum by
the eigenvalues of $(\mathcal{N}_3\mathcal{N}_3^\dagger)^{1/2}$. 
The spectrum clusters around 0, 1 and $\sqrt{2}$, as 
mentioned in Introduction and illustrated in Fig.~\ref{Fig3D}(a)
for the domain-filled state $\ket{D_B(t)}$. The bird-eye views shown in 
Figs.~\ref{Fig3D}(b,c) for the domain-filled state and 
the crystal state, respectively, demonstrate surprisingly coincidental 
movements between the contours of small singular values and the 
early-time curves of the Markov gap. 
In End Matter, we show that if the corresponding $\vec{t}'$, cf.~\eqref{eq:t1}, 
of all ETFs are tripartite, i.e., $\pi_X(\vec{t}')\neq 0,\,\forall X$, 
the singular value spectrum of $\mathcal{N}_3$ is exactly 
$\{ 0, 1, \sqrt{2}\}$ and the multiplicity of $0$ gives the
number of ETFs. Deviations from this ideal structure broaden the spectrum, reflecting a nonuique choice of of the space of bipartite fermions $V_2$, as as observed in Fig.~\ref{Fig3D}.

Thus, the contours of small singular values reflect the increasing number of ETFs. Their agreement with
the Markov gap at early times suggests an approximate proportionality, indicating that each ETF contributes comparably to the growth of the Markov gap. For long-time dynamics, however, we find discrepancy between them. This difference can be attributed to contributions from the ``holes'' inside the Fermi sea of 
bipartite fermions $\ket{\Psi_2}$, which are not captured by the small-singular-value structure of the tripartite null matrix $\mathcal{N}_3$ .

\begin{figure}
    \centering
    \includegraphics[width=\linewidth]{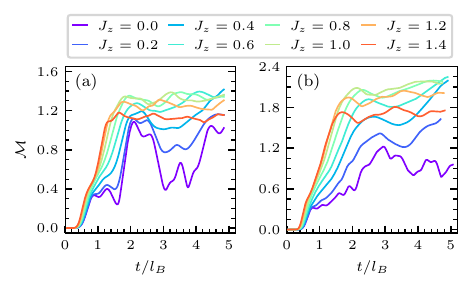}
    \caption{Dynamics of Markov gap from Neel state quenched by XXZ Hamiltonian under various strength of anisotropy $J_z$. Left and right panels are the result for system size $L=12$ and $L=18$, respectively. }
    \label{Fig5} 
\end{figure}

\emph{Spin-$1/2$ XXZ chain}---To assess the effects of interactions between fermions, we consider the spin-1/2 XXZ model
\begin{equation}
    H_2 =  H_1 + J_z \sum_{j}(n_j-\frac{1}{2})(n_{j+1}-\frac{1}{2})
    \label{Ham:B}
\end{equation}
where $n_j=c_j^\dagger c_j^{\phantom{\dagger}}$ and 
$J_z$ is a tunable parameter. 
The dynamics is simulated using the time-dependent variational principle (TDVP) method~\cite{Haegeman2011PRL,Haegeman2016PRB}, implimented with the ITensor library~\cite{ITensorA, ITensorB},
additional details are provided in Supplementary Materials.~\cite[Sec.~\ref{Supp:MPS}]{sp}. 
We initialize the system in the N{\' e}el state $ |\psi_0\rangle = |\uparrow\downarrow \cdots \uparrow\downarrow\rangle $ and set $ J_z \geq 0 $, ranging from 0.0 to 1.4 with a step of 0.2.  
The results are shown in
Fig.~\ref{Fig5} for two system sizes ($ L=12 $ and $ L=18 $).
Although finite-size effects prevent a scaling analysis, we 
still see features of $t_*$, a kink in the quick growth,
and signal of the volume-law---the saturated Markov gap at $L=18$ in Fig.~\ref{Fig5}(b) exceeds that at $L=12$ in Fig.~\ref{Fig5}(a). 
Moreover, increasing $J_z$ boosts the speed of growth at the early stage. 
As to the saturation levels,
the effect of increasing $J_z$ is positive in the
gapless phase $J_z<1.0$, but becomes 
negative in the gapped phase. This behavior mirrors the quantum criticality of the XXZ chain.

\emph{Conclusions}---In this Letter, we studied the growth of the Markov gap in 
free-fermion chains and the spin-1/2 XXZ chain. We find that the Markov gap universally exhibits volume-law scaling, 
and a sharp threshold ($t_*$) marking its onset. We proposed attainable lower and upper bounds for $t_*$,
$l_B/2v_{\max}\leq t_*\leq l_B/v_{\max}$, where $v_{\max}$ is undertstood as 
the Lieb-Robinson speed. The evolution of the Markov gap show rich fine structures with a period $t\sim L$. Depending on the
initial states, the Markov gap may grow linearly in time, oscillations, or staircase-like discrete jumps.
Beyond these fine structures, we found long-time behaviors with the time scale of $t\sim L^2$ in three examples of structured initial 
states ($\ket{C_{(3,1)}}$, $\ket{D}$, and $\ket{D_B}$). This is much longer than the time scale of 
entanglement scrambling (referred to bipartite entanglement implicitly), 
indicating that the many-body wavefunction is developing 
structures that hidden from the bipartite entanglement. 

To understand such hidden complexities, we propose the concept of essential tripartite fermion (ETF)
by exploiting the equivalent representations of an fermionic Gaussian state.
The building of ETFs, as demonstrated by the singular value spectrum of a null matrix ($\mathcal{N}$),
matches well with the profiles of the Markov gap.
We also briefly studied interacting fermions in spin-1/2 XXZ model via MPS methods. We find that the Markov gap
reaches the highest saturation at the gapless critical point. We also observe a kink
in the early-time growth of the Markov gap, which has not been seen for bipartite entanglement.
These results demonstrate that tripartite entanglement dynamics follows its own universal principles, 
distinct from and richer than those governing bipartite entanglement.

\begin{acknowledgements}
We thank Yantao Wu for stimulating discussions. Y.-X. Z. acknowledges the financial support from the National Natural Science Foundation
of China (Grant No. 12375024), and the CAS Project for
Young Scientists in Basic Research (Grant No. YSBR-100).
\end{acknowledgements}

\bibliography{References}

\section*{End Matter}

\subsection*{More About The Entanglement Structure}
Recall that states expressed as $\prod_{q=1}^n \left(\sum_{i=1}^{n}T_{q,i}d_i^\dagger\right)\ket{0}$ are equivalent to each other as long as
$T_{q,i}$ is a $n\times n$ unitary matrix. By exploiting this freedom, the bipartite fermions
between $A$ and $B$ can be determined from any basis of $F_{AB}$. Recall also that
$F_{AB}$ is embedded in $V_{AB}^{(2)}=\pi_A(F_{AB})\oplus\pi_B(F_{AB})$. Therefore, we choose
one basis $\{\vec{d}_{AB,i}\}_i$ and denote its decomposition onto $\pi_A(F_{AB})$ and $\pi_B(F_{AB})$ as
\begin{equation}
\vec{d}_{AB,i}=\vec{d}_{A,i}+\vec{d}_{B,i}.
\end{equation}
Importantly, $\{ \vec{d}_{A,i}\}_i$ are linearly independent (but not necessarily 
orthogonal to each other), otherwise by exploiting 
the unitary $T_{q,i}$ we can further reduce $F_{AB}$. For the same reason, linear
independence holds for $\{ \vec{d}_{B,i}\}_i$. Similarly, regarding to 
$F_{BC}$ and $F_{AC}$ we obtain:
\begin{equation}
    \begin{split}
        \vec{e}_{BC,j} & =\vec{e}_{B,j}+\vec{e}_{C,j}, \\
        \vec{g}_{AC,k} & =\vec{g}_{A,k}+\vec{g}_{C,k}.
    \end{split}
\end{equation}
Then, Eq.~(5) of the main text can be rewritten as follows. 
Denoting $\pi_X(\vec{t}_1)$ by $\vec{t}_{1,X}$ and $\pi_X(\vec{t}'_1)$ 
by $\vec{t}'_{1,X}$, we have the expansion 
\begin{equation} 
\label{eq:t-expansion}
\begin{aligned}
\vec{t}_{1,A} & = \sum_i\alpha_{1,Ai}\vec{d}_{A,i}+ \sum_k \gamma_{1,Ak} \vec{g}_{A,k}+ 
\vec{t}'_{1,A}\\
\vec{t}_{1,B} & = \sum_i \alpha_{1,Bi}\vec{d}_{B,i}+ \sum_j \beta_{1,Bj} \vec{e}_{B,j}+ 
\vec{t}'_{1,B},\\
\vec{t}_{1,C} & = \sum_j \beta_{1,Cj}\vec{e}_{C,i}+ \sum_k \gamma_{1,Ck} \vec{g}_{C,k}+ 
\vec{t}'_{1,C},
\end{aligned}\end{equation}
The coefficients in Eq.~\eqref{eq:t-expansion} are not all independent. For example, 
the normalization condition of $\vec{d}_{AB,i}$ reads $\abs{\vec{d}_{Ai}}^2+\abs{\vec{d}_{Bi}}^2=1$ and
the orthogonality $\vec{t}_1\perp \vec{d}_{AB,i}$ implies
$\alpha_{1,Ai}\abs{\vec{d}_{Ai}}^2+\alpha_{1,Bi}\abs{\vec{d}_{Bi}}^2=0$.

In Eq.~(6) of the main text,  $\ket{t'_1}$ is the W-type state over the 
degrees of freedom represented by
$(\vec{t}'_{1,A}, \vec{t}'_{1,B}, \vec{t}'_{1,C})$. To see $\ket{\Psi_2^{-t_1}}$, 
we rewrite $\pi_{V^{(2)}}(\vec{t}_1)$ as
\begin{equation} 
\pi_{V^{(2)}}(\vec{t}_1)= \vec{\alpha}_1 + \vec{\beta}_1+\vec{\gamma}_1
\end{equation} 
where $\vec{\alpha_1}=\sum_i\alpha_{1, Ai}\vec{d}_{A,i}+\sum_i\alpha_{1,Bi}\vec{d}_{B,i}\in V^{(2)}_{AB}$,
$\vec{\beta_1}=\sum_j \beta_{1,Bj} \vec{e}_{B,j}+\sum_j\beta_{1,Cj} \vec{e}_{C,j}\in V^{(2)}_{BC}$, 
and $\vec{\gamma_1}=\sum_k \gamma_{1,Ak} \vec{g}_{A,k}+
\sum_k\gamma_{1,Ck} \vec{g}_{C,k}\in V^{(2)}_{AC}$. 

Translating the above vector representation to the fermionic operators, it is
straightforward to calculate the W-type form of $\ket{\Psi_2^{-t_1}}$. The
expression of $\ket{\psi_{AB}^{-\alpha}}$ is
\begin{equation}
\label{eq:state-f2-not-normalized}
 \ket{\psi^{-\alpha_1}_{AB}} \equiv \alpha_1^\dagger \ket{\psi_{AB}}
 \end{equation}
Since $\ket{\psi_{AB}}$ is the full occupation of bipartite fermions between $A$ and $B$,
the effect of $\alpha_1^\dagger$ is actually to remove something. To see it clearly,
we consider an ideal situation where the projected vectors 
$\{\vec{d}_{A,i}\}_i$ are mutually orthogonal. In this case,
\begin{equation}
     \alpha_1^\dagger \ket{\psi_{AB}} = \sum_i(\alpha_{Ai}-\alpha_{Bi})\ket{\psi^{-i}_{AB}} 
\end{equation}
where $\ket{\psi_{AB}^{-i}}$ is a state where the entanglement contributed by $\vec{d}_{AB,i}$
is removed:
\begin{equation}
|\psi^{-i}_{AB}\rangle =(-1)^i\big(\prod_{m<i}\hat{d}_{AB,m}^\dagger \big)\hat{d}_{Ai}^\dagger 
\hat{d}_{Bi}^\dagger \big(\prod_{m>i}\hat{d}_{AB,m}^\dagger \big) \ket{0}.    
\end{equation}
The normalization of $|\psi^{-i}_{AB}\rangle$ is $\abs{\vec{d}_{Ai}}\abs{\vec{d}_{Bi}}$.
A few lines of derivation shows the normalization of $\ket{\psi_{AB}^{-\alpha_1}}$ is
\begin{equation} 
\braket{\psi^{-\alpha_1}_{AB}|\psi^{-\alpha_1}_{AB}}=-\sum_{i}\alpha_{Ai}\alpha_{Bi}.
\end{equation} 
The other terms in Eq.~\eqref{eq:Wt}, $\ket{\psi_{BC}^{-\beta_1}}$ and $\ket{\psi_{AC}^{-\gamma_1}}$,
are defined in the same way.

The above analysis strengthens the physical picture that tripartite entanglement
can also be generated by creating ``holes'' in the Fermi sea of bipartite fermions $\ket{\Psi_2}$.
We can translate a vector representation to a hole-operator as follows
\begin{equation} 
\vec{\alpha}_1 \rightarrow \hat{\alpha}^{-}_1
\equiv \sum_{i}(\alpha_{1,Ai}-\alpha_{1, Bi})
\hat{\Sigma}_{AB,i}^{-},
\end{equation} 
where the effect of $\hat{\Sigma}_{AB,i}^{-}$ is to map $d^\dagger_{AB,i}$ into $d_{Ai}^\dagger d_{Bi}^\dagger$. 
This formula is convenient for the case of more tripartite fermions,
where we shall need to evaluate
\begin{equation}
\hat{\alpha}_2^{-}\hat{\alpha}_1^{-}\ket{\psi_{AB}}
= \sum_{i>j}(\alpha_{1,i}\alpha_{2,j}-\alpha_{2,i}\alpha_{1,j})
\ket{\psi^{-(i,j)}_{AB}}
\end{equation} 
where $\ket{\psi^{-(i,j)}_{AB}}=
\hat{\Sigma}_{AB,j}^{-}\hat{\Sigma}_{AB,i}^{-}\ket{\psi_{AB}}$.
This formula strengthens the physical picture of viewing
the ``local projections'' of $V^{(3)}$ to $V^{(2)}$
effectively generate ``holes'' in the Fermi sea of bipartite fermions.

\subsection*{Spectrum of The Tripartite Null Matrix}

Recall that state 
$\ket{\Psi(t)}=\prod_{q}\left[\sum_{j}\Gamma_{q,j}(t)c_j^\dagger\right]\ket{0}$
can be equivalently represented by replacing the matrix $\Gamma$ with
$T\Gamma$, where  $T$ is any $n\times n$ unitary matrix.
Without loss of generality, 
let us assume that this $T$ is exactly one that extract the ETFs. 
Denote the new matrix by $\Gamma'\equiv T\Gamma=[\Gamma'_A, \Gamma'_B, \Gamma'_C]$.
The ETFs can be directly read from $\Gamma'$: If one row
is nonzero in each party, it is the vector representation of an ETF. 

Now let us assume that the nonzero rows of $\Gamma'_X$, for each $X=A, B,$ and $C$,
are linearly independent. 
This is equivalent to say $\pi_X(\vec{t}')\neq 0$ for 
every ETF and every party. ( To prove it, we need that in $\Gamma'_X$, 
the set of $\{\pi_X(\vec{t}')\}_{\text{ETF}}$ are linearly independent.
Otherwise the number of tripartite fermions can be further reduced). 
Note that if the $i$-th row of $\Gamma'_X$ is zero, the $i$-th row of
$T$ is a basis vector for the kernel of $\Gamma_X^T$.
It means that the columns of $\mathcal{N}_X$ are just the corresponding 
rows of $T$. If one row of $\Gamma'$ is nonzero only in $A$, i.e., this row
is a vector representation of a fermion localized in $A$, the
same row of $T$ will appear, as a column, twice in $\mathcal{N}_3$: For 
one time in $N_B$ and another time
in $N_C$. And if one row of $\Gamma'$ is zero only in $A$, i.e.,
it represents a bipartite fermion of $BC$,
the same row of $T$ appears for only one time. Denote the numbers of single-party, bipartite fermions, and ETFs 
by $n_1$, $n_2$ and $n_3$, respectively. The above argument 
implies that the matrix $\mathcal{N}_3\mathcal{N}_3^\dagger$ has $n_1$ diagonal 
blocks of
$\begin{pmatrix} 1 & 1 \\ 1&1 \end{pmatrix}$ and 
$n_2$ diagonal elements of 1. 

Thus, the eigenvalues of 
$(\mathcal{N}_3^\dagger \mathcal{N}_3)^{1/2}$ consist of $n_1$ multiple of $\sqrt{2}$,
$n_1$ multiple of $0$, and $n_2$ multiple of $1$.
Note that $(\mathcal{N}_3\mathcal{N}_3^\dagger )^{1/2}$ 
has the same list of nonzero eigenvalues with 
$(\mathcal{N}_3^\dagger \mathcal{N}_3)^{1/2}$, which means it has 
$n-n_1-n_2=n_3$ zero eigenvalues. This finishes the proof that
the spectrum is $\{0,1,\sqrt{2}\}$ and the multiplicity of 0 
gives the number of ETFs.

Next, we use an example to show that breaking the assumption
leads to singular values other than 0, 1, and $\sqrt{2}$. Consider 
a chain of five sites and three fermions:
\begin{equation}
\Gamma = 
\begin{tikzpicture}[baseline=(m.center)]
  \matrix (m) [matrix of math nodes,
               left delimiter={[}, right delimiter={]},
               row sep=0.5em,
               column sep=1em,
               nodes={inner sep=0pt}] 
  {
    \sqrt{0.8} & \sqrt{0.1} & \sqrt{0.1} & 0 & 0 \\[2pt]
    0          & \sqrt{0.1} & -\sqrt{0.1}& \sqrt{0.8} & 0 \\[2pt]
    0          & 0          & 0          & 0 & 1 \\
  };

  \coordinate (sep1) at ($(m-2-1.east)!0.5!(m-2-2.west)$);
  \coordinate (sep2) at ($(m-2-2.east)!0.5!(m-2-3.west)$);

  \draw[dashed, thick] (sep1 |- m.north) -- (sep1 |- m.south);
  \draw[dashed, thick] (sep2 |- m.north) -- (sep2 |- m.south);
\end{tikzpicture}
\end{equation}
where $A$, $B$, and $C$ are split by dashed vertical lines.
The Markov gap of this state is $\mathcal{M}\approx 0.18716$.
In the above matrix, the 
first row represents an ETF (denoted by $\vec{t}$). The
second row represents a bipartite 
fermion in $F_{BC}$. In this example, 
the nonzero rows of $\Gamma_B$ are not linearly independent.
Or equivalently, regarding to Eq.~\eqref{eq:t1}, now $\vec{t}'$
is not tripartite since $\pi_B(\vec{t}')=0$. Therefore,
IrTE of this state is contributed by only the ``hole''-like 
excitation. By exploiting the freedom of introducing a $T$ matrix
\begin{equation}
    T=\frac{1}{\sqrt{2}}\begin{bmatrix}
        1 & 1 & 0\\ 
        1 & -1 & 0 \\
        0 & 0 & 1
    \end{bmatrix}
\end{equation}
we can transform $\Gamma$ into 
\begin{equation*}
T\Gamma = \frac{1}{\sqrt{2}}
\begin{tikzpicture}[baseline=(m.center)]
  \matrix (m) [matrix of math nodes,
               left delimiter={[}, right delimiter={]},
               row sep=0.5em,
               column sep=1em,
               nodes={inner sep=0pt}] 
  {
    \sqrt{0.8} & 2\sqrt{0.1} & 0  & \sqrt{0.8} & 0 \\[2pt]
    \sqrt{0.8}          & 0 & 2\sqrt{0.1}& -\sqrt{0.8} & 0 \\[2pt]
    0          & 0          & 0          & 0 & \sqrt{2} \\
  };

  \coordinate (sep1) at ($(m-1-1.east)!0.5!(m-1-2.west)$);
  \coordinate (sep2) at ($(m-2-2.east)!0.7!(m-2-3.west)$);

  \draw[dashed, thick] (sep1 |- m.north) -- (sep1 |- m.south);
  \draw[dashed, thick] (sep2 |- m.north) -- (sep2 |- m.south);
\end{tikzpicture}
\end{equation*}
which corresponds to an ETF and a bipartite fermion in $F_{AC}$
and it is $\Gamma_A$ that has linearly dependent rows. 
This transformation also confirms that the choice for
the space of bipartite
fermions $V_2$ is not unique. The singular value spectrum of
the corresponding tripartite null matrix is 
\[
\bigg\{\sqrt{1-\frac{1}{\sqrt{2}}}\approx 0.54,\; 
\sqrt{1+\frac{1}{\sqrt{2}}}\approx 1.31,\; \sqrt{2} \bigg\}.
\]
Therein, $\sqrt{2}$ still marks a single-party fermion (the third row
of $\Gamma$) while 0 and 1 are replaced.

\clearpage
\thispagestyle{empty}

\setcounter{equation}{0}
\setcounter{figure}{0}
\setcounter{page}{1}

\renewcommand{\theequation}{S\arabic{equation}}
\renewcommand{\thefigure}{S\arabic{figure}}
\renewcommand{\thepage}{S\arabic{page}}

\setcounter{secnumdepth}{2} 
\renewcommand{\thefigure}{S\arabic{figure}}
\setcounter{figure}{0}
\setcounter{section}{0}
\setcounter{equation}{0}
\renewcommand{\thesection}{S-\Roman{section}}
\renewcommand{\theequation}{S\arabic{equation}}
\setcounter{enumiv}{0}
\renewcommand{\bibnumfmt}[1]{[S#1]}

\title{Supplemental Materials for \maintitle}

\date{\today}
\maketitle

\onecolumngrid

\section{Correlation matrix framework for Gaussian fermion}
\label{Supp:CM}
For Gaussian states, Wick's theorem ensures that all 
multi-point correlation functions, hence the full information, can always be expressed through
the two-point correlation functions:
\begin{equation}
\mathcal{C}_{ij} = \langle c^{\dagger}_ic_j^{\phantom{\dagger}}\rangle,
\end{equation}
where the $c_i^\dagger(c_i^{\phantom{\dagger}})$ is the fermionic creation(annihilation) operator at site $i$.  
Given an initial state represented by its correlation matrix $\mathcal{C}(0)$, its time evolution under the quench Hamiltonian $\mathcal{H}$ is governed by
\begin{equation}
\mathcal{C}(t) = {\rm exp}(i\mathcal{H}^*t)\mathcal{C}(0){\rm exp}(-i\mathcal{H}^*t),
\end{equation}  
where the superscript $*$ denotes complex conjugate.
The time-dependent von Neumann entropy of a subsystem $X$ is attainable from the reduced correlation matrix $\mathcal{C}_X(t)$ via
\begin{equation}    
S_X(t) = -{\rm Tr}\left[\mathcal{C}_X\ln \mathcal{C}_X + (1-\mathcal{C}_X)\ln (1-\mathcal{C}_X)\right].
\end{equation}
Since the canonical purification of a Gaussian state remains Gaussian, 
the reduced correlation matrix $\mathcal{C}_X(t)$ can be equivalently regarded as the correlation matrix obtained from its canonical purification in the doubled Hilbert space $X \cup X^*$. 
Specifically, after extracting the correlation matrix $\mathcal{C}_{AB}(t)$ for the joint subsystem $A\cup B$ from the full correlation matrix $\mathcal{C}(t)$, the correlation matrix of the canonically purified state on $ABA^*B^*$ takes the explicit form
\begin{equation}    
\mathcal{C}_{ABA^*B^*} = \left(\begin{matrix}    
\mathcal{C}_{AB} & \sqrt{(I-\mathcal{C}_{AB})\mathcal{C}_{AB}}\\    
\sqrt{(I-\mathcal{C}_{AB})\mathcal{C}_{AB}}& I-\mathcal{C}_{AB}    
\end{matrix}\right),
\end{equation}
where $I$ denotes the identity matrix of appropriate dimension. The reduced correlation matrix $\mathcal{C}_{AA^*}$ can then be obtained from $\mathcal{C}_{ABA^*B^*}$, enabling the direct evaluation of the reflected entropy 
\[
S_R(A:B)=S(A\cup A^*).
\]
Under this construction, the quench dynamics of the Markov gap 
$\mathcal{M}=S_R(A:B)-I({A:B})$ are accessible in fermionic Gaussian systems.

\section{Correlation Matrix for Each Kind of Initial State}
\label{Supp:Ct}

Even though we do not have a comprehensive theory to explain every
detail of the Markov gap evolution, some properties can be inferred from the
behavior of the two-point correlation matrix.
Basically,  the free fermion Hamiltonian Eq. $\left(\ref{Ham:A}\right)$ in the main text is diagonal in momentum space, $H = \sum_k \varepsilon\left(k\right) c_k^\dagger c_k$,
with the single-particle dispersion $\varepsilon(k)=- \cos(k)$ under periodic boundary condition. At early times, the Markov gap is primarily contributed by particles interfering far from the boundaries. Consequently, we neglect boundary effects for simplicity and employ plane waves to transform between real and Fourier spaces. Then, 
the time-evolved fermion operator reads
\begin{equation}
    c_x\left(t\right) = \frac{1}{\sqrt{N}}\sum_k e^{ikx - i\varepsilon(k)t}c_k\left(0\right).
\end{equation}
This form allows direct evaluation of the time-dependent correlation functions, 
\begin{equation}\label{CxytA}
    \mathcal{C}_{xy}(t) = \frac{1}{N}\!\sum_{k_1, k_2}\! e^{-i(k_1x-k_2y-\varepsilon(k_1)t+\varepsilon(k_2)t)}\langle c_{k_1}^\dagger(0) c_{k_2}^{\phantom{\dagger}}(0)\rangle.
\end{equation}
To evaluate the influences of initial states on the quench dynamics, four classes of states are considered as the initial state quenched by the free fermion chain Hamiltonian Eq.~(\ref{Ham:A}). 
In the following, we discuss the threshold emerging time $t_*$ for each initial state, 
and in particular the
$t\sim L^2$ scaling for the case of domain-filled state.

\subsection{The Crystal State}
The first is the general crystal state 
\begin{equation}	
\ket{C_{(p, q)}} = \prod_{i=1}^{L/p}\left(\prod_{j=0}^{q-1}c_{pi-j}^\dagger\right)\ket{0}.
\end{equation}
This state (and the dimer state discussed below) has a periodic pattern in the
real space. Since the spatial period is $p$ for $\ket{C_{(p,q)}}$, 
the discrete translation invariance imposes the condition that
\begin{equation}
    \mathcal{C}_{x,y}\left(t\right) = \mathcal{C}_{x+r, y+r}\left(t\right).
\end{equation}
Substituting this periodicity into Eq.~(\ref{CxytA}) yields the restriction that
\begin{equation}
    k_1 - k_2 = \frac{2\pi \nu}{p}, \quad \nu = 0, 1, \cdots, p-1.
\end{equation}
It means that although each localized fermion is uniformly distributed in the
momentum space, many-particle interference completely suppresses some momentum components.
The time-dependent correlation matrix is thus
\begin{equation}
    \mathcal{C}_{xy}(t) = \frac{1}{N}\!\sum_{k, \nu}\! e^{-i((k+Q\nu)x-ky-\varepsilon(k+Q\nu)t+\varepsilon(k)t}\langle c_{k+Q\nu}^\dagger(0) c_{k}^{\phantom{\dagger}}(0)\rangle,
    \label{CxytB}
\end{equation}
with $Q=\frac{2\pi}{p}$ the reciprocal lattice vector associated with the periodicity $p$. The dominant contributions to the correlation function arise from terms that stabilize the phase, which is determined by the equation
\begin{equation}
    (y-x) + [\varepsilon^\prime(k+Q\nu) - \varepsilon^\prime(k)]t = 0.
    \label{Stationphi}
\end{equation}
It is expanded as $t = -\frac{y-x}{\sin(k+Q\nu)-\sin(k)} = -\frac{y-x}{2 \cos(k+\frac{Q\nu}{2})\sin(\frac{Q\nu}{2})}$, which implies an effective velocity
\begin{equation}
    v_{k,\nu} = \cos(k+\frac{Q\nu}{2})\sin(\frac{Q\nu}{2}).
\end{equation}
We can define the effective light cone as the largest $v_{k,\nu}$. 
For the crystal state $\ket{C_{3,1}}$ it is thus $v_{\rm eff}=\sin(\pi/3)=\frac{\sqrt{3}}{2}v_{\max}$
(achieved by both $\nu=1$ and $2$). This is consistent with the numerical result of
$t_*=l_B/(2v_{\rm eff})$.

\subsection{The Domain-Filled State}
The threshold time $t_*$ of the domain-filled state is already discussed in the main text.
Interestingly, here we are able to gain signals of the $t\sim L^2$ long-time scaling from the
elements of the correlation matrix:
\begin{equation}
    C_{xy}(t) = \sum_{\alpha,\beta} U_{x\alpha}^*(t)U_{y\beta}(t)\langle c_\alpha^\dagger c_\beta \rangle,\label{CxyDB}
\end{equation}
where the evolution operator $ U(t) = \exp(-iHt) $ is obtained from the Heisenberg-picture time evolution.
In the initial state, the correlation matrix is diagonal
\begin{equation}
    \langle c_\alpha^\dagger c_\beta \rangle = \delta_{\alpha\beta} \quad  \text{if }   \alpha,\beta\in B,\text{ and zero otherwise}.
\end{equation}
Substituting this into Eq.~(\ref{CxyDB}) yields
\begin{equation}
    C_{xy}(t) = \sum_{\alpha\in B}U_{x\alpha}^{*}U_{y\alpha}.\label{CxyDBU}
\end{equation}
In the long time limit, the single particle propagator is asymptotically described by Fresnel wave packet
\begin{equation}
    U_{xl}(t) = \frac{1}{\sqrt{2\pi i t}}{\rm exp}\left(i\frac{(x-l)^2}{2 t}\right).
\end{equation}
For each term in the summation of Eq.~(\ref{CxyDB}), the $l^2$ cancels out and gives
\begin{equation}
    U_{xl}^*(t)U_{yl}(t) = \frac{1}{2\pi i t}{\rm exp}\left(i\frac{y^2-x^2}{2 t}\right){\rm exp}(il\frac{x-y}{ t}).
\end{equation}
Then the summation in Eq.~(\ref{CxyDBU}) involves a finite series
\begin{equation}
\begin{split}
    \sum_{l=1}^{l_B}{\rm exp}\left(i\frac{x-y}{ t}l\right) =& {\rm exp }\left(i\frac{x-y}{ t}\right)\frac{1-{\rm exp}\left(i\frac{x-y}{ t}l_B\right)}{1-{\rm exp }\left(i\frac{x-y}{ t}\right)}\\
    \approx& \frac{ t}{i(x-y)}\left({\rm exp}\left(i\frac{x-y}{ t}l_B\right)-1\right).
    \end{split}
\end{equation}
Therein, we have assumed $t\gg \abs{x-y}$ so that $e^{i(x-y)/t}\approx 1+i(x-y)/t$.
Substituting this into Eq.~(\ref{CxyDBU}) yields the asymptotic behavior in the long time limit
\begin{equation}
    C_{xy}(t)\sim \frac{1}{i2\pi(x-y)}\left({\rm exp}\left(i\frac{x-y}{t}l_B\right)-1\right){\rm exp}\left(i\frac{y^2-x^2}{2 t}\right).
\end{equation}
Now, we can see that the the phase variation of $C_{xy}(t)$  along the whole chain becomes $O(1)$ provided that $t\sim L^2$, signaling the completion of dephasing. In other word, this is the timescale when different contributions to $C_{xy}(t)$ become approximately phase aligned. As the Markov gap is a complicated nonlinear function of the correlation matrix, we expect the dephasing is crucial for the saturation of the Markov gap, resulting in the
$t\sim L^2$ long-time scaling of the latter.

\begin{figure}[tb] 
    \centering
    \includegraphics[width=1.0\linewidth]{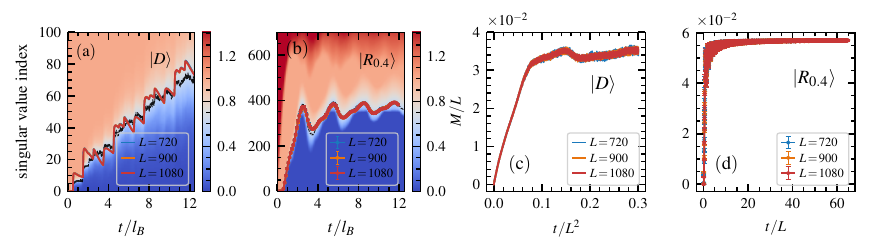}
    \caption{Panels (a) and (b) illustrate the early-time dynamics of states $|D\rangle$ and $|R_{0.4}\rangle$, respectively, across three different system sizes. For state $|D\rangle$, rescaling the time as $t/l_B$ and the vertical axis by $\mathrm{log} L$ yields a data collapse. For state $|R_{0.4}\rangle$, an excellent collapse is achieved by scaling the vertical axis by $L$. The time evolution aligns perfectly with the singular value spectrum of the tripartite null matrix. Panels (c) and (d) depict the corresponding long time dynamics. Rescaling the vertical axis by $L$ reveals a volume law in the long time limit. The different of scaling of time indicate that the state $|D\rangle$ saturates proportionally to $L^2$, whereas the state $|R_{04}\rangle$ exhibits a saturation with $L$. The dynamics of state $|R_{0.4}\rangle$ is obtained by averaging over 60 independent configurations, such small error bars indicating that the disordered self average effect works well.}
    \label{FigS2}
\end{figure}

\subsection{The Dimer State}
Next, to study the influence of short-range bipartite entanglement, we introduce the dimer state
\begin{equation}
    \left|D\right\rangle = 2^{-L/4}\prod_{p=1}^{L/2}\left(c_{2p-1}^\dagger + c_{2p}^\dagger\right)\left|0\right\rangle.
\end{equation}
In this state, each two adjacent sites form an 
independent Bell entanglement pair. This state is further 
generalized to dimer-$q$ state
\begin{equation}
     \left|D(q)\right\rangle = 2^{-L/4}\prod_{p=1}^{L/2}\left(c_{2p-1}^\dagger + \theta(p, q)c_{2p}^\dagger\right)\left|0\right\rangle.
\end{equation}
where $q$ is an integer, $\theta\left(p, q\right) = \left(-1\right)^{\lfloor p/q\rfloor \mod{2}}$. 
This construction introduces an alternating pattern of $\pm$ signs on the bonds, with a period of $2q$.

The correlation in momentum space of dimer state is given as
\begin{equation}
    \langle  c_k^\dagger c_{k^\prime}^{\phantom{\dagger}}\rangle =  \frac{\delta_{k,k'}}{2} + \frac{e^{ik'} + e^{-ik}}{4} \bigl( \delta_{|k-k'|,0} + \delta_{|k-k'|,\pi} \bigr).
\end{equation}
And the results of dimer-$q$ state is
\begin{equation}
   \langle  c_k^\dagger c_{k^\prime}^{\phantom{\dagger}}\rangle = \frac{\delta_{k,k'}}{2} 
+ \frac{1}{4q}  \sum_{p=1}^{q} \Bigl( e^{-i[(2p-1)k - 2(p-1)k']} + e^{i[2(p-1)k - (2p-1)k']} \Bigr) 
 \sum_{p=1}^{q} \Bigl( \delta_{|k-k'|,\frac{\pi(2p-1)}{2q}} + \delta_{|k-k'|,2\pi-\frac{\pi(2p-1)}{2q}} \Bigr). 
\end{equation}
Only the second term contributes the dynamics, thus the time-dependent part of the correlation matrix have the form
\begin{equation}
\mathcal{C}_{x,y}(t)\equiv \sum_{p}\sum_{k}F_{p}(k)e^{-i[kx\mp (k + \alpha_{p})y]}e^{i(\varepsilon(k) -\varepsilon(\pm (k + \alpha_{p})))t},
\end{equation}
with phase shift $\alpha_p = \frac{2p-1}{2q}\pi, p = 1,\cdots, q$, and
\begin{equation}
F_p(q) = \frac{1}{2qN}\sum_{p=1}^q \Bigl( e^{-i(2p-1)\alpha_p} + e^{i(2p-1)\alpha_p}\Bigr).
\end{equation}

The stationary phase condition contains the term
\begin{equation}
    x-y = 2t v_\text{eff}.
\end{equation}
For dimer state, $v_\text{eff} = \sin(\pi/2)=v_{\max}$. This is the example where $t_*$
reaches the lower bound we proposed in the main text.
The dimer-$q$ state hosts multi light cones. The corresponding 
velocities are 
$v_{\text{eff}, p}=2\sin\bigl(\frac{2p-1}{4q}\pi\bigr), p=1,\cdots, q$.

\subsection{Staircase Profiles of Domain-Filled State and Dimer State}
\begin{figure}[bt]
    \centering
\includegraphics[width=1.0\linewidth]{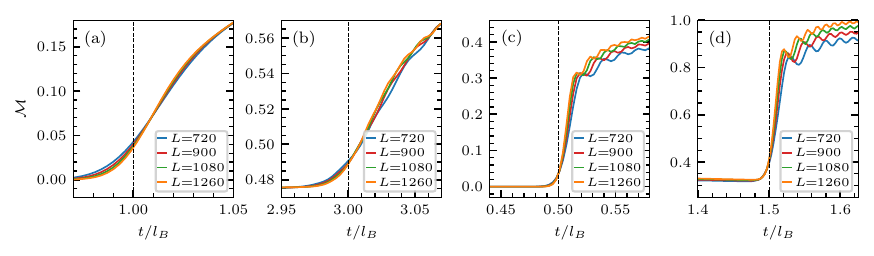}
    \caption{(a)-(b): the first two stairs of domain-wall state $|D_B\rangle$. (c)-(d): the first two stairs of the dimer state $|D\rangle$. The dashed lines marks the position of the stairs in Markov gap. In (a) and (c), the positions correspond to $t^*$, while in (b) and (d), they denote $3t^*$. }
    \label{MJumps}
\end{figure}

We have seen that the early-time dynamics of the Markov gap
show staircase like jumps for the domain-filled state $\ket{D_B}$
and the dimer state $\ket{D}$. Here we zoom in their first two 
jumps in Fig.~\ref{MJumps} for different chain size $L$.
The plots show phase-transition like profiles: The larger the size $L$, the
steeper the climbing of the Markov gap, and curves of all chain sizes 
seem to cross at the same point. 
As shown in Figs.~\ref{MJumps}(a,b), the first two jumps of the 
domain-filled state have critical points at $t/l_B\approx 1.01$ and
$3.01$, respectively. For the dimer state, the corresponding coordinates of the critical points
shown in Figs.~\ref{MJumps}(c,d) are $t/l_B\approx 0.5$ and $1.5$.
These results are consistent with our statement that in the large $L$ limit,
the threshold time $t_*$ of the domain-filled state
attains the generic upper bound $l_B/v_{\max}$ while the dimer state
attains the  lower bound $l_B/(2v_{\max})$.

A similar dynamical phase transition is seen for the GHZ entanglement
in the monitored random Clifford circuits~\cite{Xu2025prb}. Therein, if the circuit is
unitary, the critical point locates at $t_*=l_B/(2v_E)$
where $v_E$ is the so-called entanglement velocity.

\subsection{Randomly-Filled States}
At last, to account for the influence of disorder,
we consider randomly-filled states with a fixed filling factor $\delta$, represented by $\left|R_\delta\right\rangle$. The crystal state $\ket{C_{(p,q)}}$ 
belongs to a special case of $\delta=q/p$. Previously we have seen that the effectively velocity is $\max_{\nu}\sin(\nu\pi/p)$ which is less 
than $v_{\max}$ due to the many-particle interference.
But if the occupied sites in the
initial state is not arranged in a periodic pattern, we expect 
$v_{\max}$ to be the speed corresponding to the threshold time 
$t_*$. 

\begin{figure}[tb]
    \centering    \includegraphics[width=1.0\linewidth]{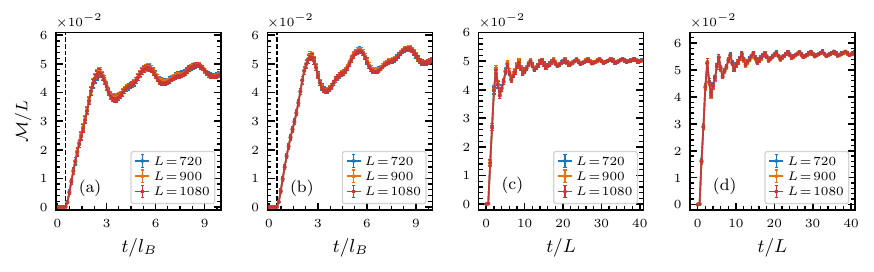}
    \caption{Early time and long time dynamics for the randomly filled state $|R_{0.2}\rangle$ in $(\mathrm{a})$, $(\mathrm{c})$, and state $|R_{0.3}\rangle$ in  $(\mathrm{b})$, $(\mathrm{d})$. The onset time is independent on the filling factor, and  saturation time scales as the system size. The saturated value is proportional to the filling factor. At a fixed filling fraction, the presented results are ensemble-averaged over 60 independent configurations.}
    \label{FigS_random}
\end{figure}

For example, we calculate the 
ensemble average of the Markov gap evolution over 60 samplings for
$\ket{R_{\delta}}$ with $\delta=0.2$ and $0.3$ and present their
early-time dynamics in Figs.~\ref{FigS_random}(a) and (b), respectively.
The threshold time is found to be $t_*=l_B/(2v_{\max})$ in both plots, as expected.
Compared with state $\ket{C_{(3,1)}}$
presented in Fig.~\ref{Fig2}(a) of the main text, 
the curves of the randomly-filled states grow almost linearly in time. We
do not observe the kink during the fast growth that presents in the
case of $\ket{C_{(3,1)}}$.

The long-time behaviors are shown in Figs~\ref{FigS_random}(c) and (d)
for $\ket{R_{\delta}}$ with $\delta=0.2$ and $0.3$, respectively.
Different from all the above states with clear structure, the long-time
behavior is featureless---the Markov gap stays at the saturation 
despite of some fine structures with period $t\sim L$.
Such behavior is seen not only in the ensemble average, but actually also
in each random sample, indicating a self-averaging effect.
The $t\sim L^2$ scaling is thus not seen here.

\section{Partitioning and Boundary Effects}
In this section, we present the Markov-gap dynamics for diverse initial states, subsystem partition ratios, and boundary conditions. As shown in Fig.~\ref{FigS2}, under open boundary conditions the onset time of a nonzero Markov gap is given by $  t = l_B / v_{\rm eff}  $, where $  l_B  $ is the length of the intermediate subsystem $  B  $ and $  v_{\rm eff}  $ is the effective quasiparticle velocity determined in the preceding section. In contrast, under periodic boundary conditions, Fig.~\ref{FigS3}, the threshold time shifts to $  t = l_m / v_{\rm eff} $, with $l_m$ the minimal subsystem length among $  A  $, $  B  $, and $C$. These findings demonstrate that genuine tripartite entanglement, as signaled by a nonzero Markov gap, emerges only after nonlocal correlations have propagated across at least one subsystem, consistent with the finite speed of information constraint imposed by the Lieb-Robinson bound. The results indicate that both the partition and boundary condition has a significant effects on the dynamics. The partitions determines the support of time-evolved fermion operators in each subsystems, it thus govern the number of ETF modes. When we adopt periodic boundary conditions, there are extra path for the information spreading and the quasiparticles has superior coherence. It is precisely for these reasons that we observe more fine structures at the periodic boundary conditions.

\begin{figure}
    \centering
    \includegraphics[width=\linewidth]{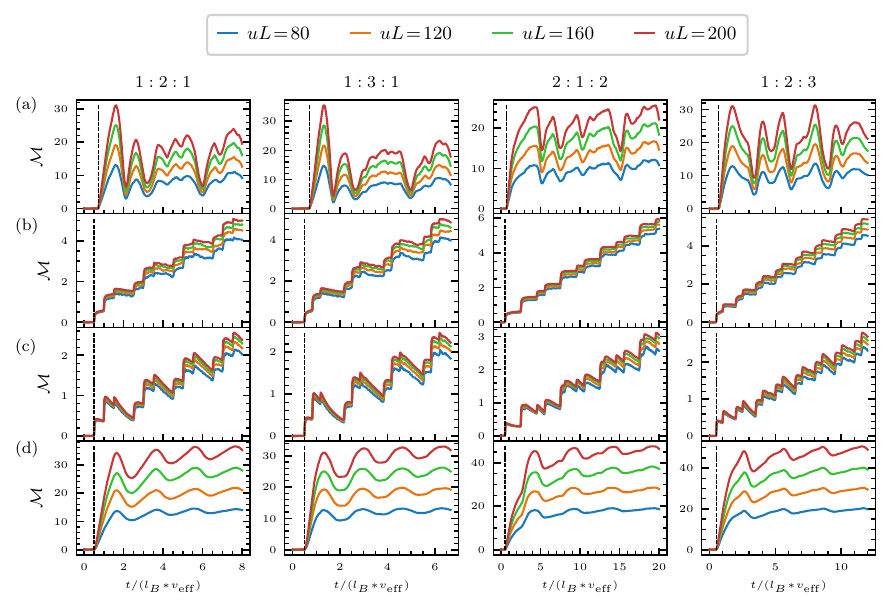}
    \caption{(a) to (d) are the Markov dynamics of state $|C_{(4, 2)}\rangle$, $|C_{(2, 1)}\rangle$, $|D\rangle$, and $|R_{0.3}\rangle$ under open boundary conditions for different system sizes, corresponding to the tripartite partitions illustrated in the top. Here, $l_B$ is the size of middle subsystem $B$. The legend indicates the unit length $uL$, and the total system size is given by $L=uL * (l_A+l_B+l_C)$ for a partition ratio $l_A:l_B:l_C$. this legend applies to all panels.}
    \label{FigS2}
\end{figure}

\begin{figure}
    \centering
    \includegraphics[width=\linewidth]{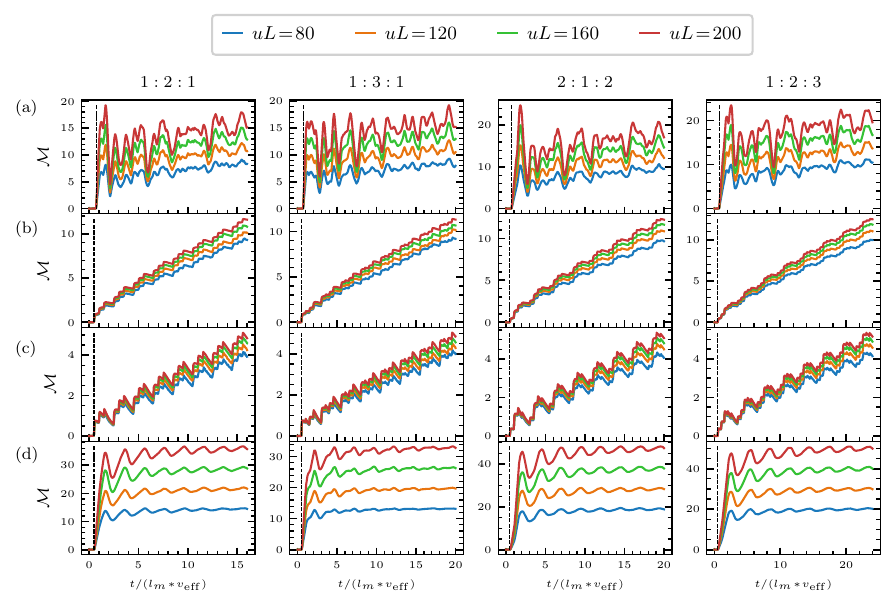}
    \caption{Panels (a)–(d) show the Markov dynamics of the states $|C_{(4,2)}\rangle$, $|C_{(2, 1)}\rangle$, $|D\rangle$, and $|R_{0.3}\rangle$ for periodic boundary conditions and the tripartite geometries displayed above. The parameter $l_m$ represents the smallest subsystem length, while the legend specifies the corresponding unit length $uL$. The total system size is given by $L=uL*(l_A+l_B+l_C)$ for the partition ratio $l_A:l_B:l_C$. All panels share the same legend.}
    \label{FigS3}
\end{figure}

\section{Matrix Product States for Markov Gap Dynamics in the XXZ Chain}
\label{Supp:MPS}
The time evolution of the XXZ spin chain is implemented via the time-dependent variational principle (TDVP). We choose a time step of $dt=0.02$ where the nearest-neighbor exchange coupling is set to $J_{xy}=2\lambda=1.0$, and represent the time-dependent wave function as a matrix product state (MPS), 
\begin{equation}
    |\psi(t)\rangle = 
     \begin{array}{c}
     \begin{tikzpicture}[scale=2.5]
        \draw[black, line width = 1.2](0.2, 0.0) --(2.7, 0.0);
        \foreach \x in {2, ..., 3}
        {
            \draw[black, line width = 1.2] (\x*0.5+0.2, +0.35) -- (\x*0.5+0.2, 0.0);
            \draw[color=black, fill = Red, rounded corners=1pt, line width = 0.9] (\x*0.5+0.2-0.13, -0.13) rectangle (\x*0.5+0.2+0.13, 0.13);
        }
        \foreach \x in {4, ..., 5}
        {
            \draw[black, line width = 1.2] (\x*0.5+0.2, +0.35) -- (\x*0.5+0.2, 0.0);
            \draw[color=black, fill = Green, rounded corners=1pt, line width = 0.9] (\x*0.5+0.2-0.13, -0.13) rectangle (\x*0.5+0.2+0.13, 0.13);
        }
        \foreach \x in {0, ..., 1}
        {
            \draw[black, line width = 1.2] (\x*0.5+0.2, +0.35) -- (\x*0.5+0.2, 0.0);
            \draw[color=black, fill = Blue, rounded corners=1pt, line width = 0.9] (\x*0.5+0.2-0.13, -0.13) rectangle (\x*0.5+0.2+0.13, 0.13);
        }
    \end{tikzpicture}
    \end{array},
\end{equation}
where we use three different colors to mark the subsystems $A$, $B$ and $C$ from left to right, respectively. To calculate the mutual information $I(A:B) = S_A + S_B - S_{C}$ and reflection entropy, we need to recast the MPS into it canonical form. For example, when the canonical center is located at the bond of MPS between $B$ and $C$
\begin{equation}
    |\psi(t)\rangle =
    \begin{array}{c}
        \begin{tikzpicture}[scale=2.5]
        \def\d{0.18} 
        \draw[black, line width = 1.0](0.2, 0.0) --(2.7, 0.0);
        \foreach \x in {4, ..., 5}
        {
            \def\r{0.5*\x};
            \draw[black, line width = 1.0, fill=Green] (\r+1.6*\d, -\d/2) -- (\r+1.6*\d, \d/2) --(\r+0.4*\d, \d/2)--(\r, 0.0)--(\r+0.4*\d, -\d/2)--cycle;
            \draw[black, line width = 0.6] (\r+\d, \d/2) -- (\r+\d, \d/2*4);
        }
        \foreach \x in {0, ..., 1}
        {
            \def\r{0.5*\x};
            \draw[black, line width = 1.0, fill=Blue] (\r, -\d/2) -- (\r, \d/2) --(\r+1.2*\d, \d/2)--(\r+1.6*\d, 0.0)--(\r+1.2*\d, -\d/2)--cycle;
            \draw[black, line width = 1.0] (\r+0.6*\d, \d/2) -- (\r+0.6*\d, \d/2*4);
        }
        \foreach \x in {2, ..., 3}
        {
            \def\r{0.5*\x};
            \draw[black, line width = 1.0, fill=Red] (\r, -\d/2) -- (\r, \d/2) --(\r+1.2*\d, \d/2)--(\r+1.6*\d, 0.0)--(\r+1.2*\d, -\d/2)--cycle;
            \draw[black, line width = 1.0] (\r+0.6*\d, \d/2) -- (\r+0.6*\d, \d/2*4);
        }
         
        \def\b{3*0.5+0.25+0.8*\d};
        \draw[rotate around={45:(\b,0)}, thick, fill = black] (\b-0.03, -0.03) rectangle (\b+0.03,0.03);
    \end{tikzpicture}
    \end{array},
\end{equation}
Then the canonical matrix $D$ located on the bond connecting $B$ and $C$ contains all the information about entanglement between $AB$ and $C$.  Here we assume that $D$ is a diagonal matrix, from whcich one can directly obtain the Schmidt decomposition of the MPS. If it is not this case,  one can reshape the MPS into this form by singular value decomposition the canonical matrix $D=U, S, V$ and absorb the $U$ and $V$ into corresponding MPS tensors. Then the entanglement entropy is available by
\begin{equation}
S_C = -\sum_{i}D_i^2\ln{D_i^2}.
\end{equation}
Similarly, $S_A$ can be obtained by move the canonical center onto the bound connecting $A$ and $B$. To calculating $S_B$, we need to obtain the reduced density martix $\rho_B$ by contracting the degrees of freedoms in $A$ and $C$
\begin{equation}
    \rho_B = 
    \begin{array}{c}
     \begin{tikzpicture}[scale=2.5]
        \def\d{0.18} 
        \def\y{-0.5}
        \def\b{3*0.5+0.25+0.8*\d};
        \draw[black, line width = 0.6, rounded corners=2pt](0.2, 0.0) --(\b, 0.0)--(\b, \y)--(0.2,
            \y)--(-0.1,\y)--(-0.1, 0,0) --(0.2, 0.0);
        \foreach \x in {0, ..., 3}
        {
            \def\r{0.5*\x};
            \draw[black, line width = 0.6, fill=Red] (\r, -\d/2) -- (\r, \d/2) --(\r+1.2*\d, \d/2)--(\r+1.6*\d, 0.0)--(\r+1.2*\d, -\d/2)--cycle;
            \draw[black, line width = 0.6, fill=Red] (\r, -\d/2+\y) -- (\r, \d/2+\y) --(\r+1.2*\d, \d/2+\y)--(\r+1.6*\d, 0.0+\y)--(\r+1.2*\d, -\d/2+\y)--cycle;
            \draw[black, line width = 0.6] (\r+0.6*\d, \d/2) -- (\r+0.6*\d, \d/2*4);
            \draw[black, line width = 0.6] (\r+0.6*\d, -\d/2+\y) -- (\r+0.6*\d, -\d/2*4+\y);
        }
        \draw[rotate around={45:(\b,0-0.15)}, thick, fill = black] (\b-0.03, -0.03-0.15) rectangle (\b+0.03,0.03-0.15);
        \draw[rotate around={45:(\b,0-0.35)}, thick, fill = black] (\b-0.03, -0.03-0.35) rectangle (\b+0.03,0.03-0.35);
    \end{tikzpicture}
    \end{array}.
\end{equation}
Then $S_B$ is calculated as 
\begin{equation}
S_B = -{\rm Tr}(\rho_B\ln{\rho_B}).
\end{equation}

As a first step to the canonical purification, we can obtain the complete MPS represent of the reduced density matrix $\rho_{AB}$
\begin{equation}
    \rho_{AB} = 
    \begin{array}{c}
          \begin{tikzpicture}[scale=2.5]
        \def\d{0.18} 
        \def\y{-0.5}
        \def\b{3*0.5+0.25+0.8*\d};
        \draw[black, line width = 0.6, rounded corners=2pt](0.2, 0.0) --(\b, 0.0)--(\b, \y)--(0.2, \y);
        \foreach \x in {0, ..., 1}
        {
            \def\r{0.5*\x};
            \draw[black, line width = 0.6, fill=Blue] (\r, -\d/2) -- (\r, \d/2) --(\r+1.2*\d, \d/2)--(\r+1.6*\d, 0.0)--(\r+1.2*\d, -\d/2)--cycle;
            \draw[black, line width = 0.6, fill=Blue] (\r, -\d/2+\y) -- (\r, \d/2+\y) --(\r+1.2*\d, \d/2+\y)--(\r+1.6*\d, 0.0+\y)--(\r+1.2*\d, -\d/2+\y)--cycle;
            \draw[black, line width = 0.6] (\r+0.6*\d, \d/2) -- (\r+0.6*\d, \d/2*4);
            \draw[black, line width = 0.6] (\r+0.6*\d, -\d/2+\y) -- (\r+0.6*\d, -\d/2*4+\y);
        }
        \foreach \x in {2, ..., 3}
        {
            \def\r{0.5*\x};
            \draw[black, line width = 0.6, fill=Red] (\r, -\d/2) -- (\r, \d/2) --(\r+1.2*\d, \d/2)--(\r+1.6*\d, 0.0)--(\r+1.2*\d, -\d/2)--cycle;
            \draw[black, line width = 0.6, fill=Red] (\r, -\d/2+\y) -- (\r, \d/2+\y) --(\r+1.2*\d, \d/2+\y)--(\r+1.6*\d, 0.0+\y)--(\r+1.2*\d, -\d/2+\y)--cycle;
            \draw[black, line width = 0.6] (\r+0.6*\d, \d/2) -- (\r+0.6*\d, \d/2*4);
            \draw[black, line width = 0.6] (\r+0.6*\d, -\d/2+\y) -- (\r+0.6*\d, -\d/2*4+\y);
        }
         
        \draw[rotate around={45:(\b,0-0.15)}, thick, fill = black] (\b-0.03, -0.03-0.15) rectangle (\b+0.03,0.03-0.15);
        \draw[rotate around={45:(\b,0-0.35)}, thick, fill = black] (\b-0.03, -0.03-0.35) rectangle (\b+0.03,0.03-0.35);
    \end{tikzpicture}
    \end{array}.
\end{equation}
From the complete representation of $\rho_{AB}$, then
\begin{equation}
\sqrt{\rho_{AB}} = 
    \begin{array}{c}
         \begin{tikzpicture}[scale=2.5]
        \def\d{0.18} 
        \def\y{-0.5}
        \def\b{3*0.5+0.25+0.8*\d};
        \draw[black, line width = 0.6, rounded corners=2pt](0.2, 0.0) --(\b, 0.0)--(\b, \y)--(0.2, \y);
        
        \foreach \x in {0, ..., 1}
        {
            \def\r{0.5*\x};
            \draw[black, line width = 0.6, fill=Blue] (\r, -\d/2) -- (\r, \d/2) --(\r+1.2*\d, \d/2)--(\r+1.6*\d, 0.0)--(\r+1.2*\d, -\d/2)--cycle;
            \draw[black, line width = 0.6, fill=Blue] (\r, -\d/2+\y) -- (\r, \d/2+\y) --(\r+1.2*\d, \d/2+\y)--(\r+1.6*\d, 0.0+\y)--(\r+1.2*\d, -\d/2+\y)--cycle;
            \draw[black, line width = 0.6] (\r+0.6*\d, \d/2) -- (\r+0.6*\d, \d/2*4);
            \draw[black, line width = 0.6] (\r+0.6*\d, -\d/2+\y) -- (\r+0.6*\d, -\d/2*4+\y);
        }
         \foreach \x in {2, ..., 3}
        {
            \def\r{0.5*\x};
            \draw[black, line width = 0.6, fill=Red] (\r, -\d/2) -- (\r, \d/2) --(\r+1.2*\d, \d/2)--(\r+1.6*\d, 0.0)--(\r+1.2*\d, -\d/2)--cycle;
            \draw[black, line width = 0.6, fill=Red] (\r, -\d/2+\y) -- (\r, \d/2+\y) --(\r+1.2*\d, \d/2+\y)--(\r+1.6*\d, 0.0+\y)--(\r+1.2*\d, -\d/2+\y)--cycle;
            \draw[black, line width = 0.6] (\r+0.6*\d, \d/2) -- (\r+0.6*\d, \d/2*4);
            \draw[black, line width = 0.6] (\r+0.6*\d, -\d/2+\y) -- (\r+0.6*\d, -\d/2*4+\y);
        }
        \draw[rotate around={45:(\b,0-0.25)}, thick, fill = black] (\b-0.03, -0.03-0.25) rectangle (\b+0.03,0.03-0.25);
    \end{tikzpicture}
    \end{array}.
\end{equation}
Immediately, we can represent $\sqrt{\rho_{AB}}$ by a MPS in the enlarged Hilbert space
\begin{equation}
    |\sqrt{\rho_{AB}}\rangle = 
    \begin{array}{c}
         \begin{tikzpicture}[scale=2.5]
        \def\d{0.18} 
        \draw[black, line width = 0.6](0.2, 0.0) --(3.5, 0.0);
        \foreach \x in {4, ..., 5}
        {
            \def\r{0.5*\x};
            \draw[black, line width = 0.6, fill=Red] (\r+1.6*\d, -\d/2) -- (\r+1.6*\d, \d/2) --(\r+0.4*\d, \d/2)--(\r, 0.0)--(\r+0.4*\d, -\d/2)--cycle;
            \draw[black, line width = 0.6] (\r+\d, \d/2) -- (\r+\d, \d/2*4);
        }
         \foreach \x in {6, ..., 7}
        {
            \def\r{0.5*\x};
            \draw[black, line width = 0.6, fill=Blue] (\r+1.6*\d, -\d/2) -- (\r+1.6*\d, \d/2) --(\r+0.4*\d, \d/2)--(\r, 0.0)--(\r+0.4*\d, -\d/2)--cycle;
            \draw[black, line width = 0.6] (\r+\d, \d/2) -- (\r+\d, \d/2*4);
        }
        \foreach \x in {0, ..., 1}
        {
            \def\r{0.5*\x};
            \draw[black, line width = 0.6, fill=Blue] (\r, -\d/2) -- (\r, \d/2) --(\r+1.2*\d, \d/2)--(\r+1.6*\d, 0.0)--(\r+1.2*\d, -\d/2)--cycle;
            \draw[black, line width = 0.6] (\r+0.6*\d, \d/2) -- (\r+0.6*\d, \d/2*4);
        }
        \foreach \x in {2, ..., 3}
        {
            \def\r{0.5*\x};
            \draw[black, line width = 0.6, fill=Red] (\r, -\d/2) -- (\r, \d/2) --(\r+1.2*\d, \d/2)--(\r+1.6*\d, 0.0)--(\r+1.2*\d, -\d/2)--cycle;
            \draw[black, line width = 0.6] (\r+0.6*\d, \d/2) -- (\r+0.6*\d, \d/2*4);
        }
         
        \def\b{3*0.5+0.25+0.8*\d};
        \draw[rotate around={45:(\b,0)}, thick, fill = black] (\b-0.03, -0.03) rectangle (\b+0.03,0.03);

    \end{tikzpicture}
    \end{array}.
\end{equation}
Here, the physical degrees of freedom in the auxiliary space is a mirror reflection of subsystem $AB$. One can rearrange the physical degrees of freedom in the auxiliary space by applying swap gate make the MPS is an representation of the state $|\sqrt{\rho_{AB}}\rangle$ in the space $H_A\otimes H_B\otimes H_A^* \otimes H_B^*$. After that, we can obtain the reduced density matrix $\rho_{AA^*}$ by contracting the physical degrees of freedom $BB^*$,
\begin{equation}
    \rho_{AA^*} = \begin{array}{c}
    \begin{tikzpicture}[scale=2.5]
    \def\d{0.18} 

    \draw[black, line width = 0.6](0.2, 0.0) --(3.5, 0.0);
    \foreach \x in {4, ..., 5}
    {
        \def\r{0.5*\x};
        \draw[black, line width = 0.6, fill=Blue] (\r+1.6*\d, -\d/2) -- (\r+1.6*\d, \d/2) --(\r+0.4*\d, \d/2)--(\r, 0.0)--(\r+0.4*\d, -\d/2)--cycle;
        \draw[black, line width = 0.6] (\r+\d, \d/2) -- (\r+\d, \d/2*4);
    }
    \foreach \x in {6, ..., 7}
    {
        \def\r{0.5*\x};
        \draw[black, line width = 0.6, fill=Red] (\r+1.6*\d, -\d/2) -- (\r+1.6*\d, \d/2) --(\r+0.4*\d, \d/2)--(\r, 0.0)--(\r+0.4*\d, -\d/2)--cycle;
        \draw[black, line width = 0.6] (\r+\d, \d/2) -- (\r+\d, \d/2*4);
    }
    \foreach \x in {0, ..., 1}
    {
        \def\r{0.5*\x};
        \draw[black, line width = 0.6, fill=Blue] (\r, -\d/2) -- (\r, \d/2) --(\r+1.2*\d, \d/2)--(\r+1.6*\d, 0.0)--(\r+1.2*\d, -\d/2)--cycle;
        \draw[black, line width = 0.6] (\r+0.6*\d, \d/2) -- (\r+0.6*\d, \d/2*4);
    }
    \foreach \x in {2, ..., 3}
    {
        \def\r{0.5*\x};
        \draw[black, line width = 0.6, fill=Red] (\r, -\d/2) -- (\r, \d/2) --(\r+1.2*\d, \d/2)--(\r+1.6*\d, 0.0)--(\r+1.2*\d, -\d/2)--cycle;
        \draw[black, line width = 0.6] (\r+0.6*\d, \d/2) -- (\r+0.6*\d, \d/2*4);
    }
     
    \def\b{3*0.5+0.25+0.8*\d};
    \draw[rotate around={45:(\b,0)}, thick, fill = black] (\b-0.03, -0.03) rectangle (\b+0.03,0.03);

    \begin{scope}[yscale=-1, shift={(0, 0.42)}]
        \draw[black, line width = 0.6](0.2, 0.0) --(3.5, 0.0);
        \foreach \x in {4, ..., 5}
        {
            \def\r{0.5*\x};
            \draw[black, line width = 0.6, fill=Blue] (\r+1.6*\d, -\d/2) -- (\r+1.6*\d, \d/2) --(\r+0.4*\d, \d/2)--(\r, 0.0)--(\r+0.4*\d, -\d/2)--cycle;
            \draw[black, line width = 0.6] (\r+\d, \d/2) -- (\r+\d, \d/2*4);
        }
        \foreach \x in {6, ..., 7}
        {
            \def\r{0.5*\x};
            \draw[black, line width = 0.6, fill=Red] (\r+1.6*\d, -\d/2) -- (\r+1.6*\d, \d/2) --(\r+0.4*\d, \d/2)--(\r, 0.0)--(\r+0.4*\d, -\d/2)--cycle;
            \draw[black, line width = 0.6] (\r+\d, \d/2) -- (\r+\d, \d/2*4);
        }
        \foreach \x in {0, ..., 1}
        {
            \def\r{0.5*\x};
            \draw[black, line width = 0.6, fill=Blue] (\r, -\d/2) -- (\r, \d/2) --(\r+1.2*\d, \d/2)--(\r+1.6*\d, 0.0)--(\r+1.2*\d, -\d/2)--cycle;
            \draw[black, line width = 0.6] (\r+0.6*\d, \d/2) -- (\r+0.6*\d, \d/2*4);
        }
        \foreach \x in {2, ..., 3}
        {
            \def\r{0.5*\x};
            \draw[black, line width = 0.6, fill=Red] (\r, -\d/2) -- (\r, \d/2) --(\r+1.2*\d, \d/2)--(\r+1.6*\d, 0.0)--(\r+1.2*\d, -\d/2)--cycle;
            \draw[black, line width = 0.6] (\r+0.6*\d, \d/2) -- (\r+0.6*\d, \d/2*4);
        }
         
        \def\b{3*0.5+0.25+0.8*\d};
        \draw[rotate around={45:(\b,0)}, thick, fill = black] (\b-0.03, -0.03) rectangle (\b+0.03,0.03);
    \end{scope}
    \draw[black, line width = 0.6,rounded corners=2pt] (1+0.6*\d, \d/2) -- (1+0.6*\d, \d/2*4+0.03)--(1+0.6*\d+0.06, \d/2*4+0.03)--(1+0.6*\d+0.06, -\d/2*4-0.45)--(1+0.6*\d, -\d/2*4-0.45)--(1.0+0.6*\d, -\d/2*4-0.4);
    \draw[black, line width = 0.6,rounded corners=2pt] (1.5+0.6*\d, \d/2) -- (1.5+0.6*\d, \d/2*4+0.03)--(1.5+0.6*\d+0.06, \d/2*4+0.03)--(1.5+0.6*\d+0.06, -\d/2*4-0.45)--(1.5+0.6*\d, -\d/2*4-0.45)--(1.5+0.6*\d, -\d/2*4-0.4);
    \draw[black, line width = 0.6,rounded corners=2pt] (3+\d, \d/2) -- (3+\d, \d/2*4+0.03)--(3+\d+0.06, \d/2*4+0.03)--(3+\d+0.06, -\d/2*4-0.45)--(3+\d, -\d/2*4-0.45)--(3+\d, -\d/2*4-0.40);
    \draw[black, line width = 0.6,rounded corners=2pt] (3.5+\d, \d/2) -- (3.5+\d, \d/2*4+0.03)--(3.5+\d+0.06, \d/2*4+0.03)--(3.5+\d+0.06, -\d/2*4-0.45)--(3.5+\d, -\d/2*4-0.45)--(3.5+\d, -\d/2*4-0.40);
\end{tikzpicture}
    \end{array},
\end{equation}
subsequently, the reflection entropy is accessible. 
\begin{figure}
    \centering
    \includegraphics[width=1.0\linewidth]{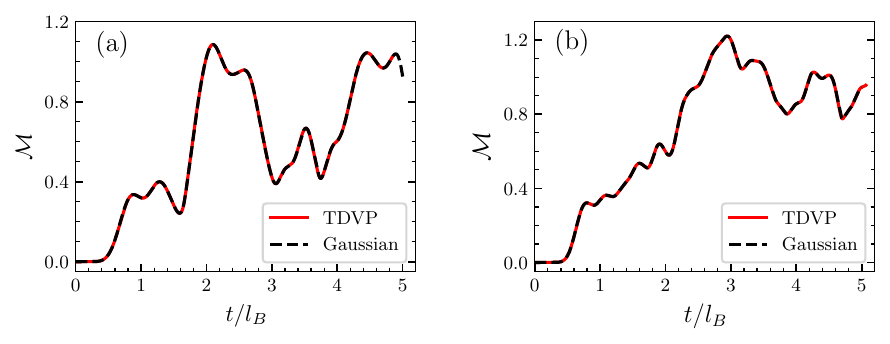}
    \caption{Benchmark of the Markov gap in the XXZ spin chain at $  J_z = 0  $. MPS calculations (red solid lines) are in excellent agreement with results obtained from the Gaussian-fermion correlation-matrix method (black dashed lines) for $L=12$ (a) and $L=18$ (b).}
    \label{Fig:S1}
\end{figure}

\end{document}